\documentclass[pdflatex,sn-basic]{sn-jnl}

\usepackage[authoryear]{natbib}
\usepackage{subfigure}
\usepackage{algorithm}
\usepackage{algorithmicx}
\usepackage{algpseudocode}
\usepackage{multirow}
\usepackage{verbatim}
\jyear{2021}%

\usepackage{color}

\theoremstyle{thmstyleone}%
\theoremstyle{thmstyletwo}%

\theoremstyle{thmstylethree}%

\begin{document}

\title[Morning commute in congested urban rail transit system]{Morning commute in congested urban rail transit system: A macroscopic model for equilibrium distribution of passenger arrivals}


\author*[1]{\fnm{Jiahua} \sur{Zhang}}\email{zhangjh@iis.u-tokyo.ac.jp}

\author[2]{\fnm{Kentaro} \sur{Wada}}\email{wadaken@sk.tsukuba.ac.jp}

\author[1]{\fnm{Takashi} \sur{Oguchi}}\email{takog.iis.u-tokyo.ac.jp}

\affil*[1]{\orgdiv{Institute of Industrial Science}, \orgname{The University of Tokyo}, \orgaddress{\street{Komaba 4-6-1}, \city{Meguro}, \postcode{1538505}, \state{Tokyo}, \country{Japan}}}

\affil[2]{\orgdiv{Faculty of Engineering, Information and Systems}, \orgname{University of Tsukuba}, \orgaddress{\street{Tennodai 1-1-1}, \city{Tsukuba}, \postcode{3058577}, \state{Ibaraki}, \country{Japan}}}


\abstract{This paper proposes a macroscopic model to describe the equilibrium distribution of passenger arrivals for the morning commute problem in a congested urban rail transit system. 
We use a macroscopic train operation sub-model developed by \cite{seo2017,seo2017b} to express the interaction between the dynamics of passengers and trains in a simplified manner while maintaining their essential physical relations. 
The equilibrium conditions of the proposed model are derived and a solution method is provided.
The characteristics of the equilibrium are then examined through analytical discussion and numerical examples.
As an application of the proposed model, we analyze a simple time-dependent timetable optimization problem with equilibrium constraints and reveal that a  “capacity increasing paradox” exists such that a higher dispatch frequency can increase the equilibrium cost. 
Furthermore, insights into the design of the timetable are obtained and the timetable influence on passengers' equilibrium travel costs are evaluated.}




\maketitle

\section{Introduction} \label{s1}
Urban rail transit, with its high capacity and punctuality, serves as a typical solution to commuters' travel demand during rush hours in most metropolises worldwide \citep{vuchic2017}. 
However, severe congestion and unexpected delays frequently degrade the travel experience of commuting by rail transit. 
In many metropolises, the congestion and delay of rail transit have brought about tremendous psychological stress to commuters and considerable economic loss to the society. 
For example, according to a report by the Ministry of Land, Infrastructure, Transport and Tourism of Japan, on an average, train delays (more than 5 min) were observed for 45 railway lines in the Tokyo metropolitan area in 11.7 days of 20 weekdays in a month, and more than half of the short delays (within 10 min) were caused by extended dwell time \citep{mlit2020}.
\cite{kariyazaki2015} estimated that in Japan, train delays resulted in social cost in excess of 1.8 billion dollars per year.

In a high-frequency operated rail transit system, when a train delay occurs because of either an accident or extended dwell time, the subsequent trains are forced to decelerate or stop between stations to maintain a safety clearance,  which is a so-called ``knock-on delay'' on the rail track \citep{carey1994knockon}.
Meanwhile, more passengers are kept waiting on the platform when trains decelerate or stop (because headways of trains are extended), which results in a longer dwell time of trains.
This is a typical vicious circle of passenger concentration and on-track congestion developed during rush hours \citep{kato2012,tirachini2013crowd,kariyazaki2015,li2017}.

To model the congestion during morning rush hours due to demand concentration, \cite{vickrey1969} proposed a departure-time choice equilibrium problem (morning commute problem) at a single bottleneck road network.
Various extensions of the basic model were proposed to address this problem in transportation planning and demand management strategies \citep[see][for a comprehensive review]{li2020}. 
However, the models for road traffic may not be readily applicable to rail transit because the mechanisms of congestion and delay differ considerably between these two systems. 
Several studies have addressed the problem in public transit systems \citep[e.g.,][]{kraus2002,tian2007,depalma2015,depalma2017,yang2018,zhang2018}.
These studies analyzed travel decisions of transit users and fare optimization issues under the assumption that the in-vehicle crowding and/or oversasrurated waiting time at stations is the primary congestion cost of traveling\footnote{In these studies, the delay of trains was not considered. Even in such a situation, a considerable waiting (queuing) time of passengers (as well as in-vehicle crowding) can occur in an oversaturated railway system \citep{shi2018,xu2019}.}.
However, no studies have addressed passengers' departure time choice behavior when considering the aforementioned vicious circle of the demand concentration and on-track congestion in a high-frequency operated rail transit system. 
Therefore, this concern should be addressed because in many metropolises, the travel time of trains between two stations (in-vehicle time) increases considerably during rush hours because of longer dwell time at intermediate stations \citep{zorn2012} and longer running time on the track \citep{kariyazaki2015,li2017}.
A comparison of the main features of this study with those of typical studies is summarized in Table \ref{tab-LR}.
\begin{table}[t]
    \caption{Comparison of typical studies on rail transit modelling with departure-time choice}
    \label{tab-LR}
    \begin{tabular}{p{15mm}p{25mm}p{10mm}p{10mm}p{10mm}p{10mm}p{13mm}}
    \toprule
    Publications & Main objectives & In-vehicle travel time & On-track congestion & In-vehicle crowding & Over-saturated waiting time & Other\quad concern \\
    \midrule
    \cite{kraus2002} & Optimal pricing, capacity and service frequency & constant & $\times$ & $\times$ & \checkmark & Operation cost \\
    \cite{tian2007} & Equilibrium property in many-to-one system & constant & $\times$ & \checkmark & $\times$ & - \\
    \cite{li2010} & Activity-based transit assignment model for timetabling problem& constant & $\times$ & \checkmark & \checkmark & - \\
    \cite{depalma2015} & Discomfort implication for scheduling and pricing & constant & $\times$ & \checkmark & $\times$ & Fare \\
    \cite{depalma2017} & Equilibrium and optimum of crowding in rail transit & constant & $\times$ & \checkmark & $\times$ & Fare \\
    \cite{yang2018} & Fare-reward scheme design & constant & $\times$ & $\times$ & \checkmark & Fare\qquad reward \\
    \cite{zhang2018} & Frequency and fares with heterogeneous users & constant & $\times$ & $\times$ & \checkmark & Operator profit \\
    \cite{zhang2020} & Modelling and optimizing congested rail transit with heterogeneous users & constant & $\times$ & \checkmark & \checkmark & Fare \\
    \cite{tang2020} & Fare scheme design with heterogeneous users & constant & $\times$ & $\times$ & \checkmark & Fare \\
    This study & Macroscopic modelling of equilibrium and timetable optimization with equilibrium constraint & time-dependent & \checkmark & $\times$ & $-^{*}$ & - \\
    \bottomrule
    \end{tabular}
    \begin{tablenotes}
    *This study only considers the undersaturated and near-saturated condition, thus the waiting time (or travel delay) is included in the travel time.
    \end{tablenotes}
\end{table}

The sophisticated microscopic modelling of congested transportation systems is typically complicated and its solution relies heavily on numerical analysis.
To describe such a complex system in a tractable manner, macroscopic fundamental diagram \citep[MFD by][]{geroliminis2007,daganzo2007} has emerged as a promising methodology in the road traffic field. 
The MFD provides simple relations among aggregate traffic variables in homogeneously congested neighborhoods, and is regarded to be useful for either economic modelling or perimeter control \citep{fosgerau2013,geroliminis2013}.
The rail transit system resembles neighborhood-size road networks such that average density or accumulation larger than some critical values degrades the throughput of the system, which corresponds to the on-track congestion of trains.
However, rail transit differs from road networks in terms of travel time because it is extended not only by the congestion of trains (vehicles), but also by the boarding and alighting passengers.
To simultaneously describe these two features of congested rail transit, \cite{seo2017,seo2017b} proposed a macroscopic and tractable train operation model (train-FD model), which enables the analysis of morning commute problem in this study. 


The purpose of this study is to develop a macroscopic model that describes the equilibrium distribution of passenger arrivals for the morning commute problem in a congested urban rail transit system. 
In this model, we use the train-FD model to express the interaction between the dynamics of passengers and trains in a simplified manner while maintaining their essential physical relations. 
We derive the equilibrium conditions of the proposed model and provide a solution method.
The characteristics of the equilibrium are then examined through both analytical discussion and numerical examples.
Finally, by applying the proposed model, we analyze a simple time-dependent timetable optimization problem with equilibrium constraints and reveal that a ``capacity increasing paradox" exists such that a higher dispatch frequency can increase the equilibrium cost. 
Furthermore, the design of timetables and their influence on passengers' equilibrium travel costs are investigated.
Owing to its simplicity and comprehensiveness, the proposed model can be a promising tool to evaluate management strategies of congested rail transit systems from both demand and supply sides.

The remainder of this paper is organized as follows.
Section \ref{s2} introduces the model for the morning commute problem in rail transit.
Section \ref{s3} derives the user equilibrium and provides a general solution method of the proposed model.
Section \ref{s4} describes the characteristics of the equilibrium through analytical discussion and numerical examples. 
Section \ref{s5} applies the proposed model to a simple time-dependent timetable optimization problem. 
Finally, conclusions and directions on future studies are discussed in Section \ref{s6}.

\section{Macroscopic model for morning commute problem in rail transit} \label{s2}
In this section, we formulate a model for the morning commute problem in rail transit. 
In Section \ref{s2-1}, we present an overview of the macroscopic train operation model proposed by \cite{seo2017,seo2017b}, which is a supply side sub-model of the proposed model. 
In Section \ref{s2-2}, we describe behavioral assumptions of users' departure time choice, which is a demand side sub-model. 


\subsection{Macroscopic rail transit operation model} \label{s2-1}
Consider a railway system on a single-line track, where stations are homogeneously located along the line. 
All trains stop at every station; thus the first-in-first-out (FIFO) service is assumed to be satisfied along the railway track. 
In the following part of this subsection, we first present the microscopic operation assumptions\footnote{Throughout this paper, we do not consider the costs and revenue of the transit operator/agency. 
Thus, we treat the train operation exogenously except for Section \ref{s5} in which an optimal timetable setting is discussed from the passenger's perspective.} on {\it passenger boarding} and {\it train cruising} to obtain a macroscopic model.

Passenger boarding behavior is described using a queuing model \citep{wada2012}. 
Specifically, the train dwell time $t_{b}$ at each station is expressed as follows:
\begin{align} \label{etb}
	t_{b} = t_{b0} + a_{p}h/\mu,
\end{align}
where $t_{b0}$ is the buffer time, including the time required for door opening and closing, $\mu$ is the maximum flow rate of passenger boarding, $a_{p}$ is the passengers' arrival rate at the platform, and $h$ is the headway of two succeeding trains. 
Note that we assume that passengers can always board the next train.

The cruising behavior of trains is assumed to be described by Newell's simplified car-following model \citep{newell2002}. 
In this model, a vehicle either travels at its desired speed or follows the preceding vehicle while maintaining safety clearance\footnote{This assumption is more appropriate for moving block rather than fixed block railway signaling system.}. 
Specifically, the position of train $n$ at time $t$ is described as follows:
\begin{align} \label{exm}
	x_{n}(t) = \min\{x_{n}(t - \tau) + v_{f}\tau, x_{n-1}(t - \tau) - \delta\},
\end{align}
where $n-1$ refers to the preceding train of train $n$, $\tau$ is the reaction time of the train, and $\delta$ is the minimum spacing. 
The first term represents the free-flow regime, where the train cruises at its desired speed $v_{f}$. 
The second term represents the congested regime, where the train decreases its speed to maintain minimum spacing.

A train fundamental diagram (train-FD), $q = Q(k, a_{p})$ describes the steady-state relation among train flow $q$ ($q=1/h$), train density $k$, and passenger arrival rate $a_{p}$ in a homogeneously congested transit system. 
Specifically, based on the operating principles described in Eqs. \eqref{etb} and \eqref{exm}, train-FD can be analytically expressed as follows \citep[see][for a derivation]{seo2017,seo2017b}: 
\begin{align} \label{eFD}
	Q(k,a_{p})=\left\{
	\begin{aligned}
	&\frac{kl - a_{p}/\mu}{t_{b0} + l/v_{f}}, & \mbox{if }k<k^*(a_{p}), \\
	&-\frac{\delta l}{(l-\delta)t_{b0} + \tau l}(k-k^*(a_{p}))+q^*(a_{p}), & \mbox{if }k\ge k^*(a_{p}),
	\end{aligned}
	\right.
\end{align}
where $l$ is the (average) distance between adjacent stations, and $q^*(a_{p})$ and $k^*(a_{p})$ are the critical train flow and train density, respectively: 
\begin{align} 
	& q^*(a_{p}) = \frac{1 - a_{p}/\mu}{t_{b0} + \delta /v_{f} + \tau},\label{eFD-b1}\\
	& k^*(a_{p}) = \frac{(1 - a_{p}/\mu)(t_{b0} + l/v_{f})}{(t_{b0} + \delta /v_{f} + \tau)l} + \frac{a_{p}}{\mu l}.\label{eFD-b2}
\end{align}

The train-FD was inspired by the MFD for road networks, and they are similar as follows.
First, they both describe the traffic states in a homogeneously congested area using system-wide aggregate variables. Second, they both exhibit unimodal relations between the density (accumulation) and flow (throughput) of the system, which yields two regimes: the free-flow and congested regimes. 
An essential difference between the train-FD and MFD is that the train-FD has an additional dimension of passenger flow. 
Introducing this new dimension enables the simplified modeling of rail transit operations in which passenger concentration is considered\footnote{Empirical investigations of the train-FD can be found in \cite{fukuda2019} and \cite{zhangjh2020}.}.



\begin{figure}[t]
    \centering
    \includegraphics[width=0.8\textwidth]{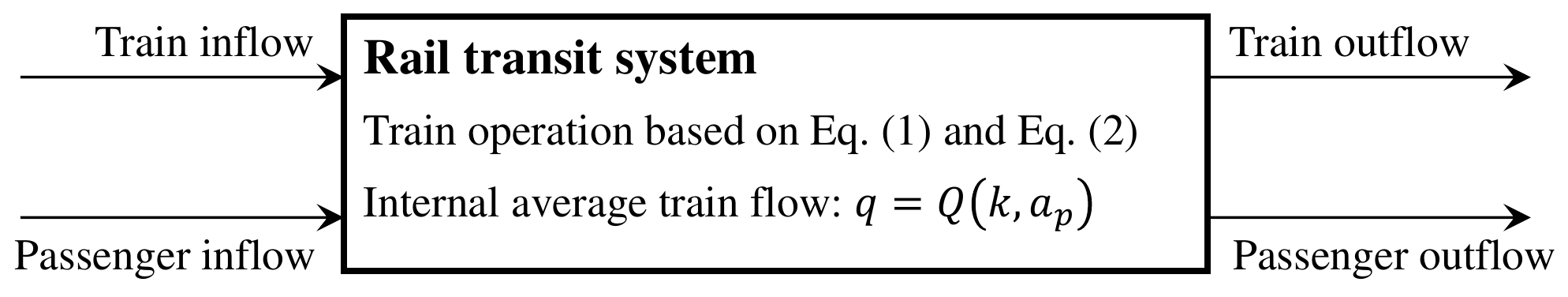}
    \caption{Rail transit system as an input-output system.}
    \label{fig-inputoutput}
\end{figure}

To describe the rail transit system behavior when the demand (i.e., passenger flow) and supply (i.e., train density) change dynamically, we consider the system as an input-output system with the train-FD, as illustrated in Fig. \ref{fig-inputoutput}. 
Two types of inputs exist: train inflow (equivalent to timetable information) and passenger inflow (i.e., passenger arrival rate $a_{p}$). 
Accordingly, two outputs are considered: train outflow and passenger outflow.
Within the system, the trains operate based on the rules in Eqs. \eqref{etb} and \eqref{exm}, whereas passenger arrival in the system is based on their assessment of the travel cost introduced in the next subsection.  
As in existing MFD applications for morning commute problems \citep[e.g., ][]{geroliminis2009,geroliminis2013,fosgerau2015}, it is expected that this simplified model can provide insight into the time-dependent characteristics of the rail transit system despite its inability to capture spatial dynamics or heterogeneity within the system.


\subsection{Passenger travel cost} \label{s2-2}
Consider a fixed number $N_{p}$ of passengers that use the train system during the morning rush period. 
The length of their trip in the system is common for all passengers and is denoted by $L$. 
Passengers choose their departure time from home to minimize their travel costs.
The travel time from leaving home to arriving at the nearest station for any passenger is assumed to be constant; thus, without a loss of generality, it is set to zero.
We also assume that the departure time from the system is the arrival time at the destination (i.e., the workplace).
For clarity, if not particularly indicated, we refer to ``passenger/train departure'' as the departure (or exit) from the rail transit system.

The travel cost (TC) is assumed to consist of the travel delay cost (TDC) in the train system and schedule delay cost (SDC)\footnote{This paper does not consider dynamic pricing. 
Thus, a constant fare is excluded from the cost function.}.
Specifically, the TC of a passenger $i$ departing from the system at time $t$ is defined as follows:
\begin{align} \label{eTTC}
	TC(t, t_i^*) = \alpha \left(T(t) - T_0 \right) + s(t, t_i^*),
\end{align}
where $t_i^*$ is the desired departure time, $\alpha$ is the time value for a travel delay, $T(t)$ is the travel time for a passenger departing from the rail transit system at time $t$, $T_0$ is the minimum travel time before the morning rush starts, and $s(t, t_i^*)$ is the schedule delay cost. 
Here, we employ the following piecewise linear schedule delay cost function $s(t, t_i^*)$ that has been widely used in previous studies \citep[e.g.,][]{hendrickson1981,tian2007,geroliminis2009,depalma2017,yang2018}. 
\begin{align} \label{eSD}
	s(t, t_i^*)=\left\{
	\begin{aligned}
	&\beta (t_i^* - t), & \mbox{if }t<t_i^*, \\
	&\gamma (t - t_i^*), & \mbox{if }t\ge t_i^*,
	\end{aligned}
	\right.
\end{align}
where $\beta$ and $\gamma$ are the values of time for earliness and lateness, respectively. 
For simplicity, we assume that all passengers have the same cost parameters $\alpha$, $\beta$ and $\gamma$.
We specify the desired departure time distribution in a later section.

\begin{figure} [t]
    \centering
    \includegraphics[width=0.5\textwidth]{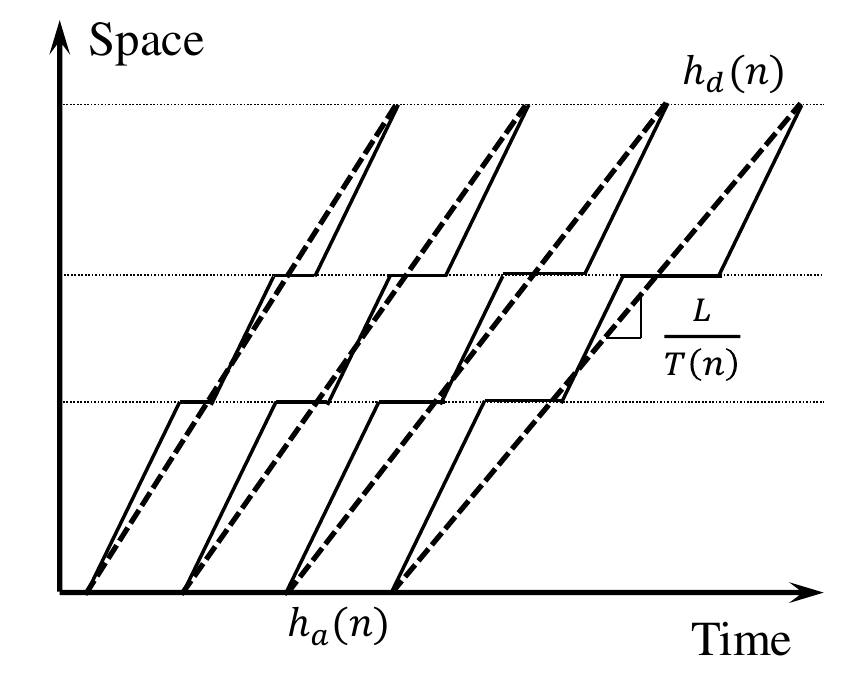}
    \caption{Example of trajectories of trains and definition of variables.} \label{fig-def} 
\end{figure}

The travel time for a passenger departing from the rail transit system at time $t$ is equal to that of a train departing from the system at the same time. 
Let $n$ be train number\footnote{Because we treat trains as a continuum or fluid, the number of trains can be a non-integer value.} departing from the system at time $t$, and $T(n) \ (= T(t))$ be its travel time.
The service (or average traveling) speed of train $n$ is $L/T(n)$. 
We denote the headway of train $n$ when arriving at the system by $h_{a}(n)$ and that when departing from the system by $h_{d}(n)$.
If we approximate the time-space trajectories of the trains as straight lines whose slopes are their service speeds (see Fig. \ref{fig-def}), the average spacing of train $n$, $\overline{s}(n)$, can be defined as follows:
\begin{align} \label{eq-snhn}
    & \overline{s}(n) \equiv \cfrac{L}{T(n)} \ \overline{h}(n)\\
    & \text{where} \quad \overline{h}(n) = \cfrac{h_{a}(n) + h_{d}(n)}{2}. \notag
\end{align}
This expression is consistent with Edie's generalized definition of traffic variables \citep{edie1963}. 

Here, we introduce the main assumption in this study: the (average) train flow $q(n) = 1/\overline{h}(n)$ and (average) train density $k(n) = 1/\overline{s}(n)$ of the system with respect to train $n$ satisfy the train-FD, that is,
\begin{align}
    & q(n) = Q(k(n), a_{p}(n)) \quad \Leftrightarrow \quad  \frac{1}{\overline{h}(n)} = Q\left(\cfrac{1}{\overline{s}(n)}, a_{p}(n)\right). \label{eq:assumption}
\end{align}
where $a_{p}(n)$ is the average passenger arrival rate for train $n$ at the stations along the line. 
If the system is in a steady state, the relation in \eqref{eq:assumption} must hold.
Therefore, the relation in \eqref{eq:assumption} approximately holds if the (average) values of the state variables vary gradually. 

This assumption enables us to link the time-dependent (more precisely, train-dependent\footnote{Because the cumulative number of train departures from the system at time $t$, $D(t)$, is an increasing function of $t$, a one-to-one correspondence between the number of trains and their departure time exists, that is, $n = D(t) \ \Leftrightarrow t = D^{-1}(n)$.}) passenger demand $\{a_{p}(n)\}$ to travel time $\{T(n)\}$ in a simplified manner while maintaining their essential physical relationships.
Specifically, as illustrated in the next section, the train traffic state variables are determined by the departure-time choice equilibrium conditions first, and the equilibrium passenger arrival rates $\{a_{p}(n)\}$ can then be estimated using the relation in \eqref{eq:assumption}.  
Notably, Eq. \eqref{eq:assumption} does not represent macroscopic train system dynamics. 
The train system dynamics using the train-FD (i.e., an exit-function model) can be found in \cite{seo2017,seo2017b}. 


\section{User equilibrium} \label{s3}
Under the setting described in the previous section, the user equilibrium is defined as the state in which no transit user can reduce his/her travel cost by changing his/her departure time from the system unilaterally. 
In this section, we first derive the equilibrium conditions. 
Next, we present a solution method of the proposed model.

\subsection{Equilibrium conditions}
Because each passenger $i$ selects his/her departure time $t_{i}$ from the system to minimize the travel cost at equilibrium, the following condition is satisfied at time $t = t_{i}$: 
\begin{align} \label{eEq}
	\frac{\partial TC(t_{i}, t_i^*)}{\partial t} = \alpha \cfrac{\mathrm{d}T(t_{i})}{\mathrm{d}t} + \cfrac{\partial s(t_{i}, t_{i}^{*})}{\partial t} = 0.
\end{align}
The derivative of the travel time $T(t)$ is obtained by substituting Eqs. \eqref{eTTC} and \eqref{eSD} into Eq. \eqref{eEq} as follows:
\begin{align} \label{edTT}
	\frac{dT(t_i)}{dt}=\left\{
	\begin{aligned}
	&\beta/\alpha, & \mbox{if }t_i<t_i^*, \\
	&-\gamma/\alpha, & \mbox{if }t_i\ge t_i^*.
	\end{aligned}
	\right.
\end{align}
Furthermore, with the first-in-first-work assumption \citep{daganzo1985}, the travel time $T(t)$ is maximized when the schedule delay is zero (we refer to this time as $t_{m}$).
Consequently, the travel time $T(t)$ under equilibrium is expressed as follows:
\begin{align} \label{eTT}
	T(t)=
	\begin{cases}
	T_0 + \frac{\beta}{\alpha}(t - t_0)=T_{e}(t),& \mbox{if }t_0\leq t<t_{m}, \\
	T_0 + \frac{\beta}{\alpha}(t_{m} - t_0) - \frac{\gamma}{\alpha}(t - t_{m})=T_{l}(t),& \mbox{if }t_{m}\leq t\leq t_{ed},
	\end{cases}. 
\end{align}
where $t_0$ and $t_{ed}$ represent the start and end of the morning rush period, respectively. 

As mentioned in the previous section, the equilibrium passenger arrivals are estimated using the train traffic state variables (i.e., $T(n), h_{a}(n), h_{d}(n)$). 
Because we have already specified the travel time under the equilibrium $T(n) = T(D^{-1}(n))$ through Eq. \eqref{eTT}, the headways for all dispatched trains are derived as follows. 
Let $A(t)$ be the cumulative number of train arrivals at the system at time $t$. 
Next, the FIFO condition is expressed as $D(t) = A(t - T(t))$ or in its derivative form as follows:
\begin{align} \label{edt}
	d(t) = a\left( t - T(t) \right)\left( 1 - \frac{dT(t)}{dt} \right), 
\end{align}
where $a(t)$ and $d(t)$ are the inflow and outflow of the trains, respectively. 
Because $A(t)$ is the given information (i.e., timetable), $D(t)$ can be obtained from this FIFO condition.
From the definition, $h_{a}(n)$ and $h_{d}(n)$ are derived as follows:
\begin{align} \label{eheadwaydefi}
   h_{a}(n) = \frac{\mathrm{d}(t-T(t))}{\mathrm{d}n} = \frac{1}{a\left( t - T(t) \right)},\quad h_{d}(n) = \frac{\mathrm{d}t}{\mathrm{d}n} = \frac{1}{d(t)}.
\end{align}
We can then calculate the average headway $\overline{h}(n)$ and spacing $\overline{s}(n)$ from these variables. 

Now, we can estimate the passenger arrivals under equilibrium. 
For a given train density, the train-FD provides a one-to-one correspondence between the train and passenger flows, that is, $q = \hat{Q}(a_{p} \mid k) = Q(a_{p}, k)$. 
Therefore, from our main assumption \eqref{eq:assumption}, we have the following expression:
\begin{align} \label{eq-ap}
    a_{p}(n) = \hat{Q}^{-1}\left(\cfrac{1}{\overline{h}(n)} \bigm\vert\cfrac{1}{\overline{s}(n)}\right)
\end{align}
where we use the following inverse function $a_{p} = \hat{Q}^{-1}(q \mid k)$.

\begin{figure}[t]
\centering
\subfigure[Case WT1]{\includegraphics[scale=0.39]{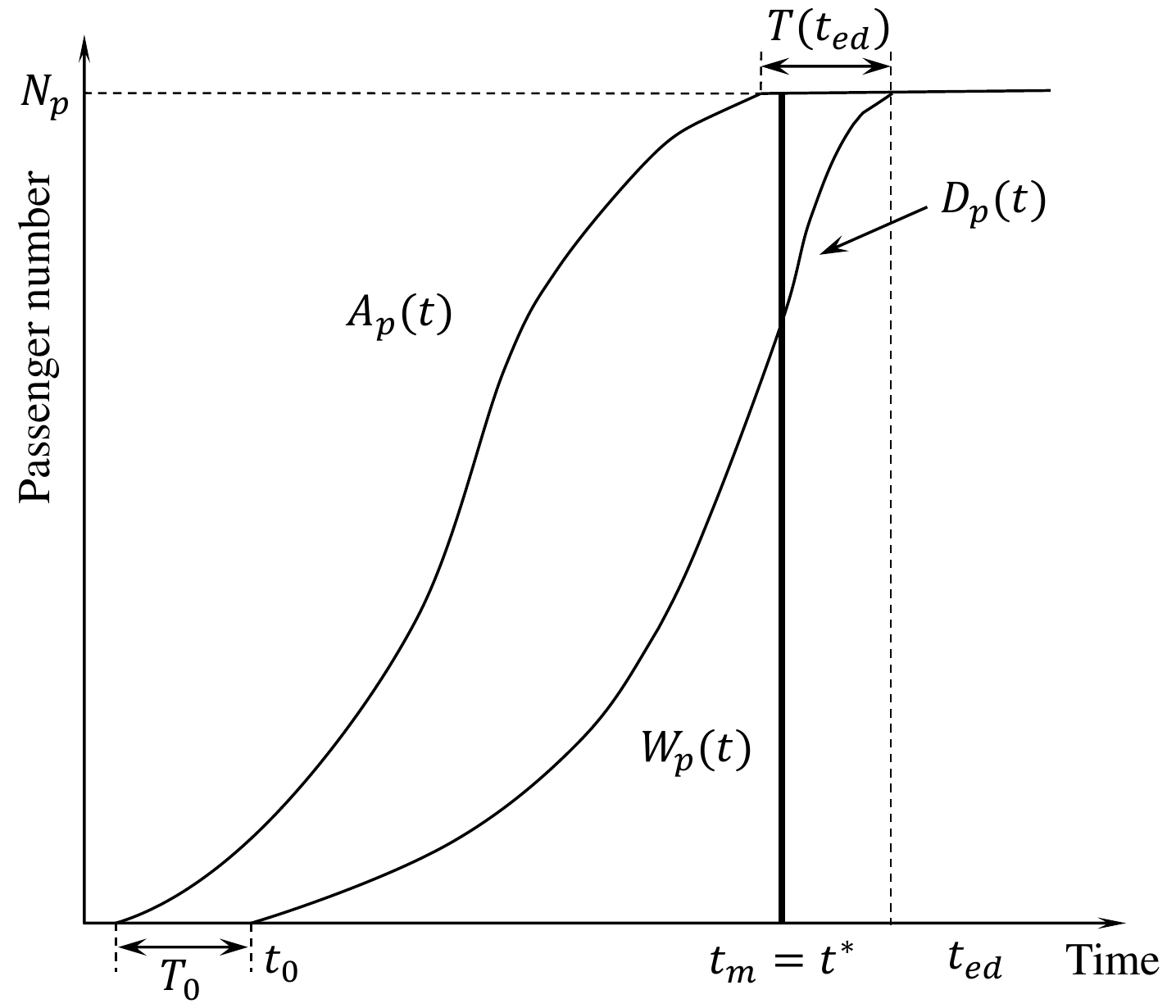}}\hspace{5pt}
\subfigure[Case WT2]{\includegraphics[scale=0.39]{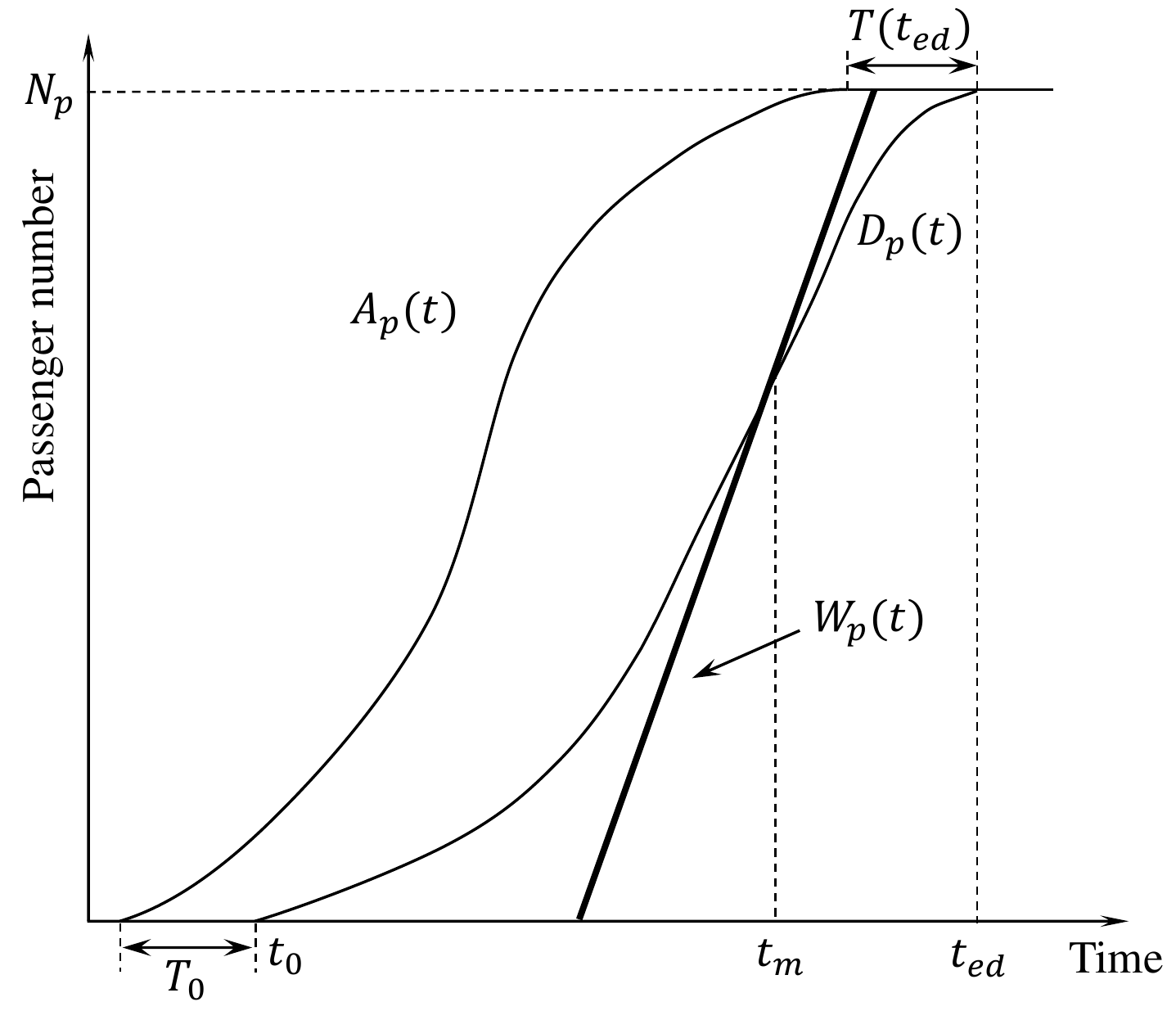}}
\caption{An illustration of cumulative curves of passengers.} \label{fig-cumpaxCD} 
\end{figure}

To obtain a complete equilibrium solution (i.e., to determine $t_{0}$, $t_{m}$ and $t_{ed}$), the desired departure time distribution should be specified. 
In this study, we consider two types of distributions: Cases WT1 and WT2. 
For Case WT1,  a fixed number $N_{p}$ of passengers has a common desired departure time $t^{*}$ (or work start time).
For Case WT2, the cumulative number of passengers who want to depart by time $t$ is expressed by using a Z-shaped function, $W_{p}(t)$, with $N_{p}$ passengers and a positive constant slope (i.e., demand rate)  \citep[e.g.,][]{gonzales2012}. An illustration of the cumulative curves of passengers for these two cases is displayed in Fig. \ref{fig-cumpaxCD}.


For Case WT1, the first condition is $t_{m} = t^{*}$. 
As the second condition, the last user experiences only the schedule delay cost, that is:
\begin{align} \label{ttted}
	T(t_{ed}) = T_0 = \frac{L}{l} \left( t_{b0} + l/v_{f} \right).
\end{align}
The last condition is the conservation of the number of users expressed as follows:
\begin{align}\label{demandcons}
   D_{p}(t_{ed}) = \int_{D(t_{0})}^{D(t_{ed})}a_{p}(n)\overline{h}(n)\mathrm{d}n = N_{p} 
\end{align}
where $D_{p}(t)$ is the cumulative number of passengers departing from the system at time $t$, and $D_{p}(t_{0}) = 0$. 
By solving the last two conditions simultaneously, $t_{0}$ and $t_{ed}$ are determined.  

For Case WT2, we assume that a unique time instant $t_{m}$ exists when the schedule delay becomes zero, as in the standard morning commute problem for road traffic \citep{smith1984,daganzo1985}. We then have the following expression:
\begin{align} \label{Wthat}
	D_{p}(t_{m}) = W_{p}(t_{m}).
\end{align}
By solving the three conditions \eqref{ttted}, \eqref{demandcons} and \eqref{Wthat} simultaneously, $t_{0}$, $t_{m}$ and $t_{ed}$ are determined. 

A solution method for Case WT1 is presented in Algorithm \ref{alg_caseS}, where $\Delta t$ is the step size of time, $\Delta n$ is the discrete unit of the train, and $\epsilon_{p}$ is the tolerance of error in the number of passengers.
The solution method for Case WT2 is similar to Algorithm \ref{alg_caseS}, that is, another step is simply added to calculate $t_{m}$ that satisfies Eq. \eqref{Wthat} after line 1. 
Note that an equilibrium solution may not exist (i.e., the solution method can produce a physically infeasible result). 
This problem is addressed in the next section. 

\begin{algorithm} [tb]
\caption{Solution to Case WT1} \label{alg_caseS}
\begin{algorithmic}[1]
\Require
Operational parameters, $l$, $L$, $t_{b0}$, $\mu$, $v_{f}$, $\delta$, $\tau$; cost parameters, $\alpha$, $\beta$, $\gamma$, $t^*$; train inflow, $a(t)$, and total travel demand, $N_{p}$.
\Ensure
Train flow $q(n)$, train density $k(n)$, and passenger arrival rate $a_{p}(n)$.
\State Set an initial $t_0$.
\State Calculate $T(t)$ and $t_{ed}$ by Eqs. \eqref{eTT} and \eqref{ttted}. \label{alg2}
\State Calculate $d(t)$ using Eq. \eqref{edt}.
\State Calculate $a_{p}(n)$ by Eq. \eqref{eq-ap}, together with Eqs. \eqref{eq-snhn} and \eqref{eheadwaydefi}.  \label{alg4}
\State Calculate the LHS $-$ RHS of the discrete version of Eq. \eqref{demandcons} (with unit $\Delta n$), denoted as an $error$. \label{alg6}
\If {$error < -\epsilon_{p}$,}
\State $t_0 = t_0 - \Delta t$, repeat lines \ref{alg2}-\ref{alg6}.
\ElsIf {$error > \epsilon_{p}$,}
\State $t_0 = t_0 + \Delta t$, repeat lines \ref{alg2}-\ref{alg6}.
\Else
\State Calculation converges, $t_0$ and $t_{ed}$ are determined.
\EndIf
\State Outputs are obtained from line \ref{alg4} when the calculation converges.
\end{algorithmic}
\end{algorithm}

\section{Characteristics of equilibrium} \label{s4}
\subsection{Analytical discussion} \label{s4-1}
In this subsection, the characteristics of equilibrium are examined analytically.
Specifically, we derive the analytical solution of the proposed model for some train operation patterns and discuss the equilibrium flow and cost structure.
For clarity, we only consider Case WT1 hereinafter.
First, by using Eqs. \eqref{edTT}, \eqref{eTT}, \eqref{edt} and \eqref{eheadwaydefi}, the average flow $q(n)$ and average density $k(n)$ for each train $n$ ($t=D^{-1}(n)$) can be expressed as follows:
\begin{align}
    & q(n) = \frac{1}{\bar{h}(n)} = 
    \begin{cases}
    \frac{2(\alpha-\beta)}{2\alpha-\beta}a\left(t-T(t)\right)=\zeta_1 a\left(t-T(t)\right),& \mbox{if }t_0\leq t<t_{m}, \\
    \frac{2(\alpha+\gamma)}{2\alpha+\gamma}a\left(t-T(t)\right)=\zeta_2 a\left(t-T(t)\right),& \mbox{if }t_{m}\leq t\leq t_{ed},
    \end{cases}\label{qndiscuss}\\
    & k(n) = \frac{1}{\bar{s}(n)} = \frac{T(n)}{\bar{h}(n)L}=
    \begin{cases}
    \frac{\zeta_1}{L}a\left(t-T(t)\right)T_{e}(t),& \mbox{if }t_0\leq t<t_{m}, \\
    \frac{\zeta_2}{L}a\left(t-T(t)\right)T_{l}(t),& \mbox{if }t_{m}\leq t\leq t_{ed}.
    \end{cases}\label{kndiscuss}
\end{align}
It can be understood from Eq. \eqref{qndiscuss} that the average flow or average headway of a train under equilibrium is determined only by the inflow $a(t-T(t))$ when this train enters the railway system and two time-value parameters, namely $\zeta_1<1$ and $\zeta_2>1$.
To guarantee that the flow calculated using Eq. \eqref{qndiscuss} is always positive, one feasibility condition of the time value should be satisfied:
\begin{align} \label{const1e}
	\alpha > \beta.
\end{align}
This condition is consistent with that of the equilibrium models for road traffic \citep[e.g.,][]{hendrickson1981,arnott1990economics}.

When the train inflow is a constant, all the trains exiting the system before $t_{m}$ exhibit the same average flow smaller than the inflow, and all the trains exiting the system after $t_{m}$ exhibit the same average flow larger than the inflow.
The average density of a train under equilibrium is the product of the time value parameter, train inflow, and piece-wise linear travel time.
Thus, when the inflow is constant, the density linearly increases until $t_{m}$ and subsequently linearly decreases until the end of the equilibrium period.

The average passenger arrival rate $a_{p}(n)$ for train $n$ can also be expressed explicitly using Eq. \eqref{eq-ap} and the aforementioned flow and density.
For the free-flow regime, it can be written as follows:
\begin{align} \label{apn_ff}
    a_{p}(n) = 
    \begin{cases}
    \mu\zeta_1\frac{l}{L} a(t-T(t))\frac{\beta}{\alpha}(t-t_0), & \mbox{if }t_0\leq t<t_{m}, \\
    \mu\zeta_2\frac{l}{L} a(t-T(t))\left[\frac{\beta}{\alpha}(t_{m}-t_0)-\frac{\gamma}{\alpha}(t-t_{m})\right], & \mbox{if }t_{m}\leq t\leq t_{ed}.
    \end{cases}
\end{align}
For the congested regime, let $\eta=(l-\delta)t_{b0}+\tau l$, $a_{p}(n)$ can be written as:
\begin{align} \label{apn_cong}
    a_{p}(n) = 
    \begin{cases}
    \frac{\mu}{l-\delta}\left[l-a(t-T(t))\left(\delta\zeta_1\frac{l}{L}T_{e}(t) - \eta\zeta_1\right)\right], & \mbox{if }t_0\leq t<t_{m}, \\
    \frac{\mu}{l-\delta}\left[l-a(t-T(t))\left(\delta\zeta_2\frac{l}{L}T_{l}(t) - \eta\zeta_2\right)\right], & \mbox{if }t_{m}\leq t\leq t_{ed}.
    \end{cases}
\end{align}
From Eq. \eqref{apn_ff} and Eq. \eqref{apn_cong}, we see that $a_{p}(n)$ relates to time-dependent $a(t-T(t))$, which leads to the difficulty in deriving an analytical solution. 
Therefore, we consider train inflow as a given constant, that is, $a(t)=a_{c}$, hereinafter.
Because the train-FD has two regimes, the evolution of railway dynamics under equilibrium relates to the combination of the two regimes in different ways.
All the possible patterns are FF, FCF, and FCCF, where F represents the free-flow regime and C represents the congested regime.
The sequences of F and C indicate the order of occurrence of these two regimes.

\subsubsection{Pattern FF}
When all dispatched trains under equilibrium operate in the free-flow regime of train-FD, it indicates that only Eq. \eqref{apn_ff} should considered. Thus, $a_{p}(n)$ first linearly increases from 0 and maximizes at $t_{m}$; subsequently it linearly decreases to 0 at the end of the equilibrium period.
The analytical solution can be obtained explicitly for the FF. 
First, the conservation law in Eq. \eqref{demandcons} is rewritten as follows by using Eq. \eqref{eheadwaydefi}, Eq. \eqref{edTT}, Eq. \eqref{edt} and $n=D(t)$:
\begin{align} \label{conservationdt}
    \int_{D(t_{0})}^{D(t_{ed})}a_{p}(n)\overline{h}(n)\mathrm{d}n &= \int_{t_{0}}^{t_{ed}}a_{p}(n)\frac{h_{a}(n)+h_{d}(n)}{2}d(t)\mathrm{d}t\\ \nonumber
    &=\left(1-\frac{\beta}{2\alpha}\right)\int_{t_{0}}^{t_{m}}a_{p}(n)\mathrm{d}t + \left(1+\frac{\gamma}{2\alpha}\right)\int_{t_{m}}^{t_{ed}}a_{p}(n)\mathrm{d}t\\ \nonumber
    &= N_{p}.
\end{align}
Then, by substituting Eq. \eqref{apn_ff} into Eq. \eqref{conservationdt}, the equilibrium cost $TC^{e}=\beta(t_{m}-t_0)=\gamma(t_{ed}-t_{m})$, is obtained as follows:
\begin{align} \label{TC_ffsolution}
    TC^{e}=\sqrt{\frac{2\alpha LN_{p}}{\mu la_{c}\left(\frac{1}{\beta}+\frac{1}{\gamma}\right)}}.
\end{align}
The equilibrium cost $TC^{e}$ increases with an increase in $N_{p}$, $\alpha$, $\beta$ and $\gamma$, and decreases with the increase in $\mu$ and $a_{c}$.
This means that under free-flow operation, lower total travel demand, higher passenger boarding rate and higher dispatch frequency could reduce the equilibrium cost.

The occurrence condition for FF is given by constraining $k(n)$ around $t_{m}$ smaller than the critical density $k^{*}$ of the train-FD. Specifically, this condition can be expressed as follows:
\begin{align} \label{ffkas}
    \left\{
    \begin{aligned}
    \frac{\zeta_1}{L}a_{c}\left[T_0 + \frac{\beta}{\alpha}(t_{m}-t_{0})\right]\leq \frac{1}{l}\left[1+\zeta_1 a_{c}\left(\frac{l-\delta}{v_{f}}-\tau\right)\right],\\
    \frac{\zeta_2}{L}a_{c}\left[T_0 + \frac{\beta}{\alpha}(t_{m}-t_{0})\right]\leq \frac{1}{l}\left[1+\zeta_2 a_{c}\left(\frac{l-\delta}{v_{f}}-\tau\right)\right].\\
    \end{aligned}
    \right.
\end{align}
Because $\zeta_2>1>\zeta_1$, the first line in Eq. \eqref{ffkas} is satisfied as long as the second line is satisfied. Thus, substituting Eq. \eqref{TC_ffsolution} into the second line, we obtain the following expression:
\begin{align} \label{Npbound1}
    TC^{e} \leq \frac{\alpha L\left[1+\zeta_2 a_{c}\left(\frac{l-\delta}{v_{f}}-\tau\right)\right]}{\zeta_2 l a_{c}} - \alpha T_{0} = TC^{\text{FF}}.
\end{align}
We denote the unique and maximum passenger demand $N_{p}$, satisfying Eq. \eqref{Npbound1} as $N_{p}^{\text{FF}}$.  
\subsubsection{Pattern FCF}
When $N_{p}$ exceeds $N_{p}^{\text{FF}}$, the second line in Eq. \eqref{ffkas} first breaks, whereas the first line still holds.
This condition indicates that all the trains exiting the railway system before $t_{m}$ operate in the free-flow regime of train-FD, whereas a portion of trains exiting after $t_{m}$ are forced to operate in the congested regime of train-FD.
This operation pattern is referred to as FCF.
In FCF, $a_{p}(n)$ may exhibit one or two peaks during the rush hour depending on whether the value of the first line in Eq. \eqref{apn_ff} is smaller than that of the second line in Eq. \eqref{apn_cong} at time $t_{m}$.
For the one-peak case, the peak occurs when the train state switches from the congested regime to the free-flow regime at time $t > t_{m}$.
For the two-peak case, another peak occurs at time $t_{m}$.
By considering the second line in Eq. \eqref{apn_cong} together with Eq. \eqref{apn_ff} in the conservation law \eqref{conservationdt}, the analytical solution for the FCF is obtained as follows:
\begin{align} \label{TC_ffcongsolution}
    & TC^{e} = \beta\left(\frac{-R + \sqrt{R^2 - 4U(S-N_{p})}}{2U}\right), 
\intertext{where}
    & R = \frac{\mu\beta}{(l-\delta)\gamma}\left(1+\frac{\gamma}{2\alpha}\right)\left[l-\eta\zeta_2 a_{c}-\delta\zeta_2 \frac{l}{L}a_{c}T_{0}\right], \notag\\
    & S = \mu\left(1+\frac{\gamma}{2\alpha}\right)\frac{TC^{\text{FF}}}{\gamma}\left[\frac{l}{2\alpha L}\zeta_2 a_{c}TC^{\text{FF}}+\frac{\delta}{l-\delta}\zeta_2\frac{l}{2\alpha L}a_{c}(2\alpha T_{0}+TC^{\text{FF}})\right]\notag\\
    &\quad + \mu\left(1+\frac{\gamma}{2\alpha}\right)\frac{TC^{\text{FF}}}{\gamma}\left(\frac{\eta}{l-\delta}\zeta_2 a_{c} - \frac{l}{l-\delta}\right),\notag\\
    & U = \frac{\mu l}{2L}a_{c}\left[\frac{\beta}{\alpha}\left(1-\frac{\beta}{\alpha}\right) - \frac{\delta\beta^2}{(l-\delta)\alpha\gamma}\left(1+\frac{\gamma}{\alpha}\right)\right]. \notag
\end{align}
The sensitivity of $TC^e$ to $N_{p}$ is also obtained as follows:
\begin{align} \label{parTCparNp}
    \frac{\partial TC^{e}}{\partial N_{p}} = \frac{\beta}{\sqrt{R^2+4UN_{p}-4US}}.
\end{align}
As long as the setting of parameters ensures that $R^2+4UN_{p}-4US\ge0$, $TC^e$ for FCF monotonically increases with $N_{p}$, and its sensitivity to $N_{p}$ decreases with an increase in $N_{p}$ when $U>0$ and increases with an increase in $N_{p}$ when $U<0$.
However, the sensitivities of $TC^e$ to time values $\alpha$, $\beta$, and $\gamma$ are tedious to derive from Eq. \eqref{TC_ffcongsolution}.
Therefore, we use numerical experiments to examine this issue in the next subsection.

As in FF, according to Eq. \eqref{ffkas}, the occurrence condition for the FCF is as follows:
\begin{align} \label{ffcongbound}
    TC^{\text{FF}} < TC^{e}\leq \frac{\alpha L\left[1+\zeta_1 a_{c}\left(\frac{l-\delta}{v_{f}}-\tau\right)\right]}{\zeta_1 l a_{c}} - \alpha T_{0} = TC^{\text{FCF}},
\end{align}
We denote the maximum demand $N_{p}$ satisfying Eq. \eqref{ffcongbound} as $N_{p}^{\text{FCF}}$.
\subsubsection{Pattern FCCF}
When $N_{p}$ further exceeds $N_{p}^{\text{FCF}}$, the operation of one portion of trains exiting the railway system before $t_{m}$ also enters the congested regime of train-FD.
Immediately after $t_{m}$, trains still operate in the congested regime because both flow and density suddenly increase after $t_{m}$.
Therefore, a sudden decrease in $a_{p}(n)$ occurs, as can be understood from Eq. \eqref{apn_cong}.
Next, the density continuously decreases after $t_{m}$ until the operation returns to free-flow again in the end.
This operation pattern is referred to as FCCF.
In this pattern, two peaks of $a_{p}(n)$ appearing separately for trains exiting before and after $t_{m}$ exist.
We skip the tedious derivation of the analytical solution for this pattern.

The physical feasibility condition (for FCCF) is that $a_{p}(n)$ calculated from the second line in Eq. \eqref{apn_cong} should be non-negative when $N_{p}$ or equivalently $TC^{e}$ is sufficiently large, that is 
\begin{align}
    a_{p}\left(D(t_{m})\right)\ge 0 \quad \Leftrightarrow \quad TC^{e}\leq \frac{\alpha L(l-\eta\zeta_2 a_{c})}{\delta\zeta_2 l a_{c}} - \alpha T_{0} = TC^{\text{FCCF}}.
\end{align}
Meanwhile, FCCF occurs when $TC^{e}>TC^{\text{FCF}}$ (or equivalently $N_{p}>N_{p}^{\text{FCF}}$).
Combining this condition and the feasibility condition, one may find that FCCF exists when $TC^{\text{FCF}}<TC^{e}\leq TC^{\text{FCCF}}$.
Note that $TC^{\text{FCCF}}>TC^{\text{FCF}}$ holds if
\begin{align}
    \alpha L\left[\left(\frac{1}{\delta\zeta_2}-\frac{1}{l\zeta_1}\right)\frac{1}{a_{c}} - \left(\frac{1}{\delta}-\frac{1}{l}\right)\left(t_{b0} + \delta/v_{f} + \tau\right)\right]>0.
\end{align}
Because $\zeta_2>1>\zeta_1>0$ and generally $l>\delta$, we can further find the following relationships: 
\begin{align}
    \frac{1}{\delta\zeta_2}-\frac{1}{l\zeta_1} < \frac{1}{\delta}-\frac{1}{l},\quad \frac{1}{a_{c}} > t_{b0} + \delta/v_{f} + \tau = \frac{1}{q^{*}(0)}.
\end{align}
Thus, for a given rail transit system (i.e., $l$, $\delta$, $t_{b0}$, $v_{f}$, and $\tau$ are given), FCCF is more likely to exist when passenger time value parameters $\zeta_1$ and $\zeta_2$ are close to 1 and/or the dispatch headway is much larger than the minimum headway.

Based on the discussion in this subsection, we conclude that, depending on the total travel demand and operational parameters of the railway system, the operation of the railway system under equilibrium can be completely in the free-flow regime or in the combination of free-flow and congested regimes.
The analytical solutions of the equilibrium are derived for two of three patterns under a constant train inflow condition. 
The feasibility and occurrence conditions of the three patterns are also determined.


\subsection{Numerical examples}
In this subsection, the characteristics of the equilibrium introduced in the previous subsection are examined through several numerical examples.
The basic parameter settings are listed in Table \ref{t-pasetting}. 
For simplicity, the train inflow $a(t)$ was set as a constant. 
For Case WT1, the common desired departure time was set to $240$ min, whereas for Case WT2, the slope of the Z-shaped function was set to $w_{p} = 30000$ pax/h, and the time period for the increase in $W_{p}(t)$ was $[210, 270]$ min. 

\begin{table}[t]
  \centering
  \caption{Parameter settings for numerical example.}
    {\begin{tabular}{cccc}
    \toprule
    \textbf{Parameter} & \textbf{Value} & \textbf{Parameter} & \textbf{Value} \\
    \midrule
    $l$     & 1.2 km & $\alpha$ & 20 $\$/$h \\
    $L$     & 18 km & $\beta$  & 8 $\$/$h \\
    $v_{f}$    & 40 km/h & $\gamma$ & 25 $\$/$h \\
    $t_{b0}$   & 20 sec & $t^*$    & 240 min \\
    $\mu$    & 36000 pax/h & $a(t)$  & 12 tr/h \\
    $\delta$ & 0.4 km & $w_{p}$   & 30000 pax/h \\
    $\tau$   & 1.0 min & $N_{p}$    & 30000 pax \\
    $\Delta t$   & 1.0 min & $\epsilon_{p}$    & 100 pax \\
    $\Delta n$   & 1 tr & \\
    \bottomrule
    \end{tabular}}%
  \label{t-pasetting}%
\end{table}%

\begin{figure}[t]
\centering
\subfigure[Case WT1]{\includegraphics[scale=0.39]{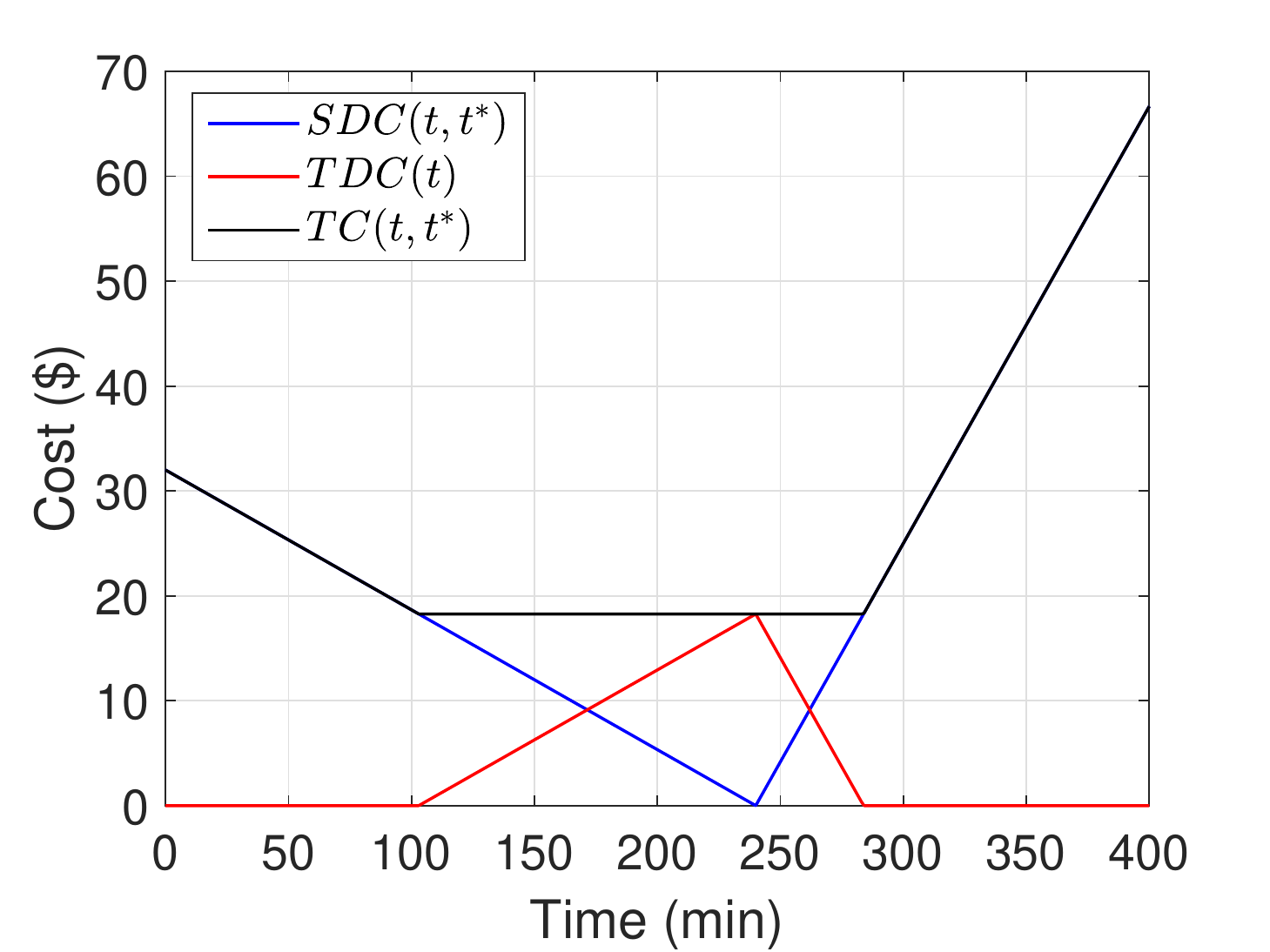}}\hspace{5pt}
\subfigure[Case WT2 with $t_i^* < t_{m}$]{\includegraphics[scale=0.39]{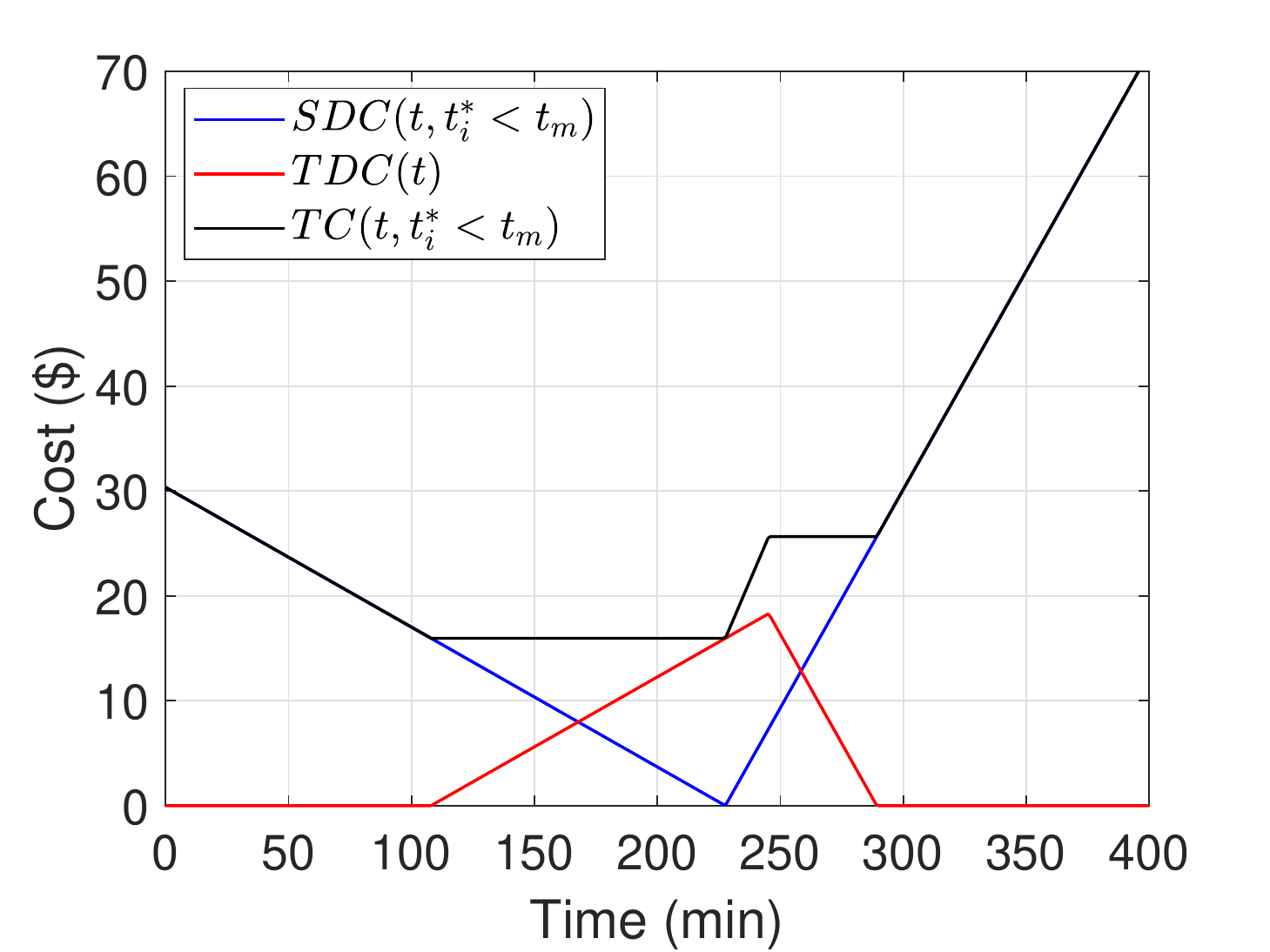}}
\caption{Travel cost for two cases.} \label{fig-ttc}
\end{figure}

\begin{figure}[t]
\centering
\subfigure[Cumulative number of trains]{\includegraphics[scale=0.39]{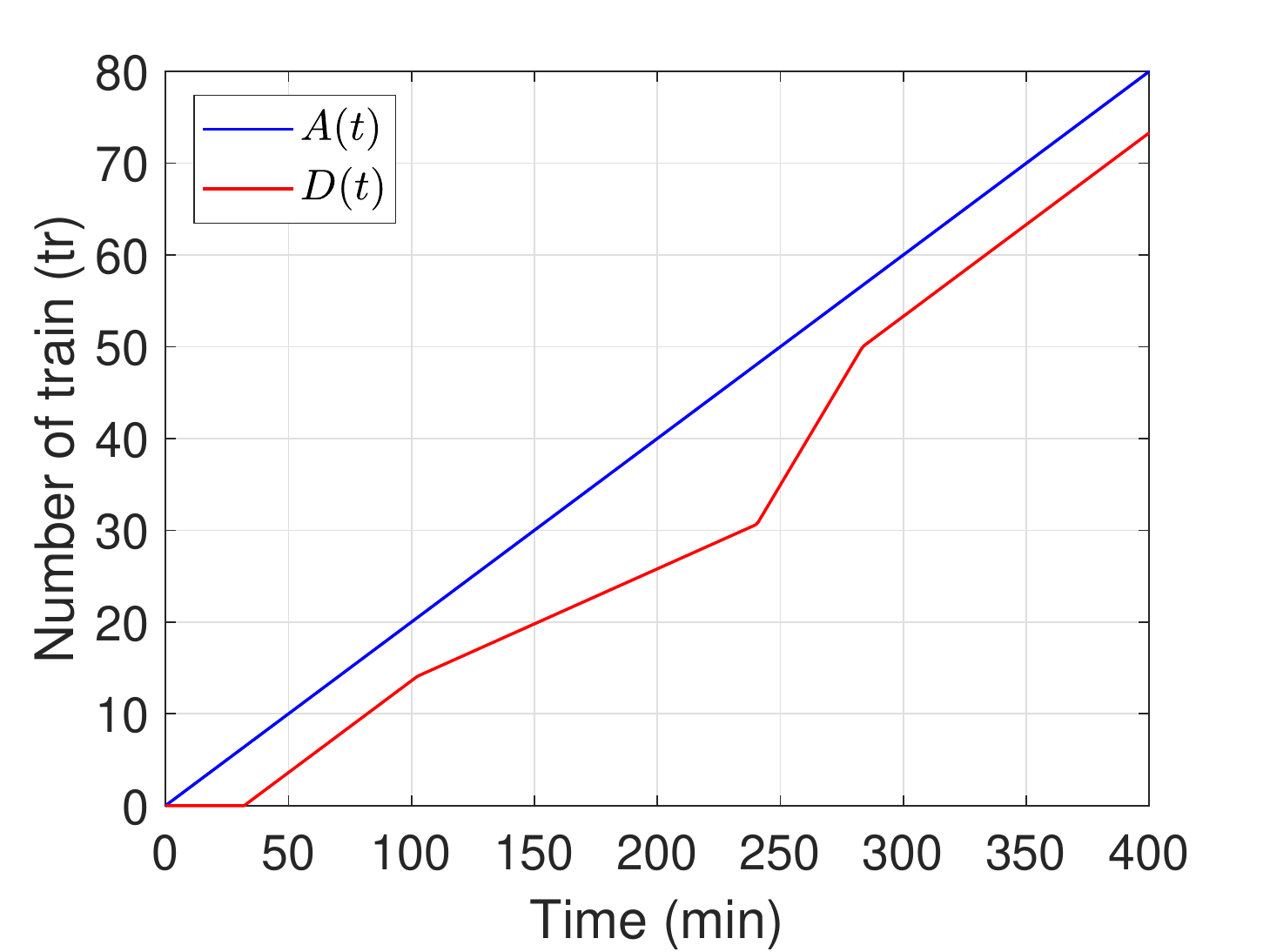}}\hspace{5pt}
\subfigure[Cumulative number of passengers]{\includegraphics[scale=0.39]{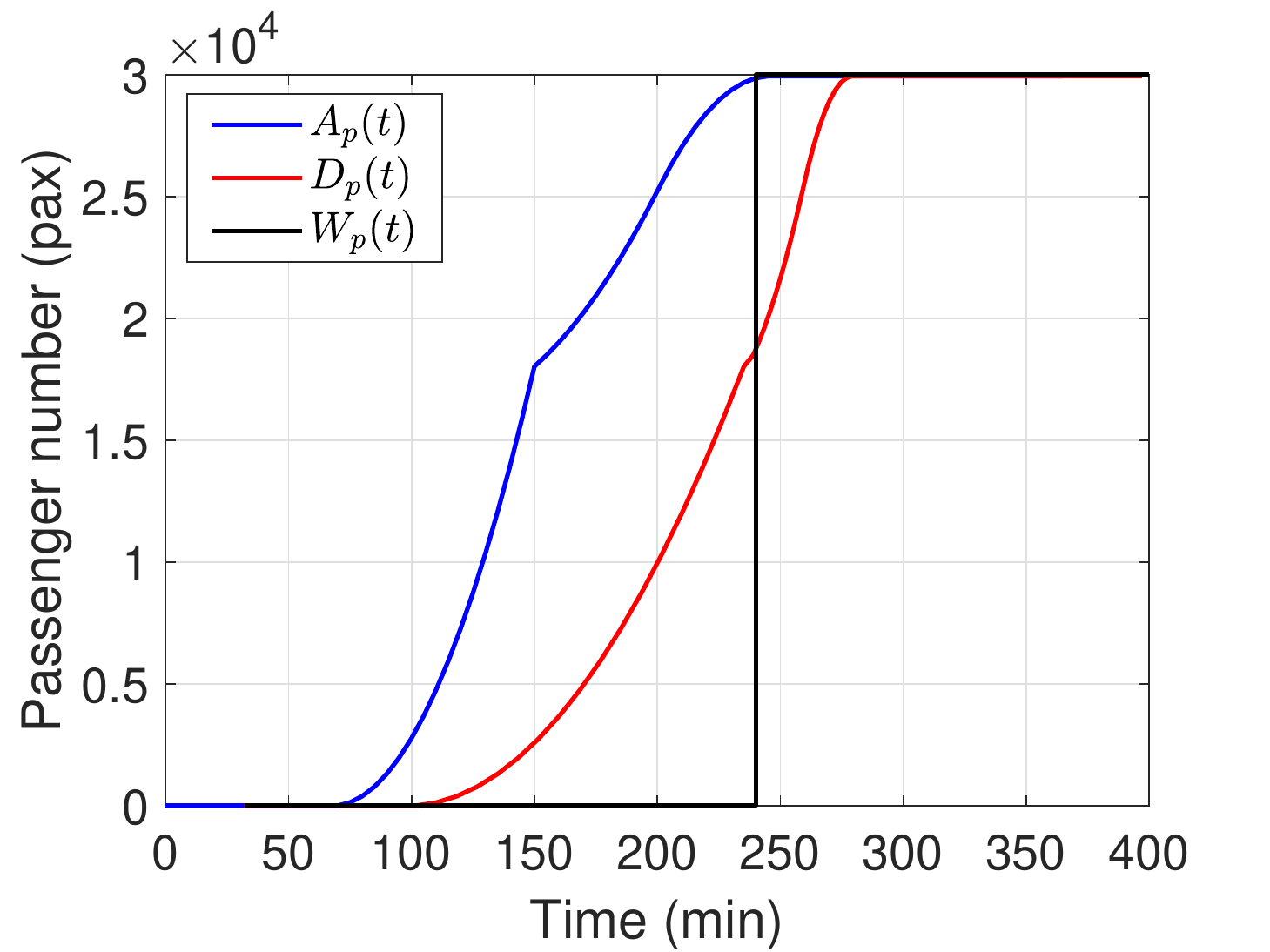}}
\caption{Dynamics of the rail transit system.} \label{fig-traindynamics} 
\end{figure}

We first present the costs for Cases WT1 and WT2 in Fig. \ref{fig-ttc}, which reveals that the cost pattern is the same as the standard morning commute problem for road traffic with a piecewise linear schedule delay cost function.
The dynamics of rail transit and passengers for Case WT1 are displayed in Fig. \ref{fig-traindynamics}. 
Fig. \ref{fig-traindynamics}(a) shows the cumulative arrival and departure curves of the trains.
Fig. \ref{fig-traindynamics}(b) shows the cumulative arrival and departure curves of the passengers. 
The train and passenger dynamics for Case WT2 are almost the same as those in these figures. 
Fig. \ref{fig-traindynamics}(a) reveals that $D(t)$ first deviates from $A(t)$ during $[t_0, t^*]$ and again approaches $A(t)$ during $[t^*,  t_{ed}]$. 
This train system behavior leads to an equilibrium in the travel cost. 



\begin{figure}[t]
    \centering
    \includegraphics[width=0.6\textwidth]{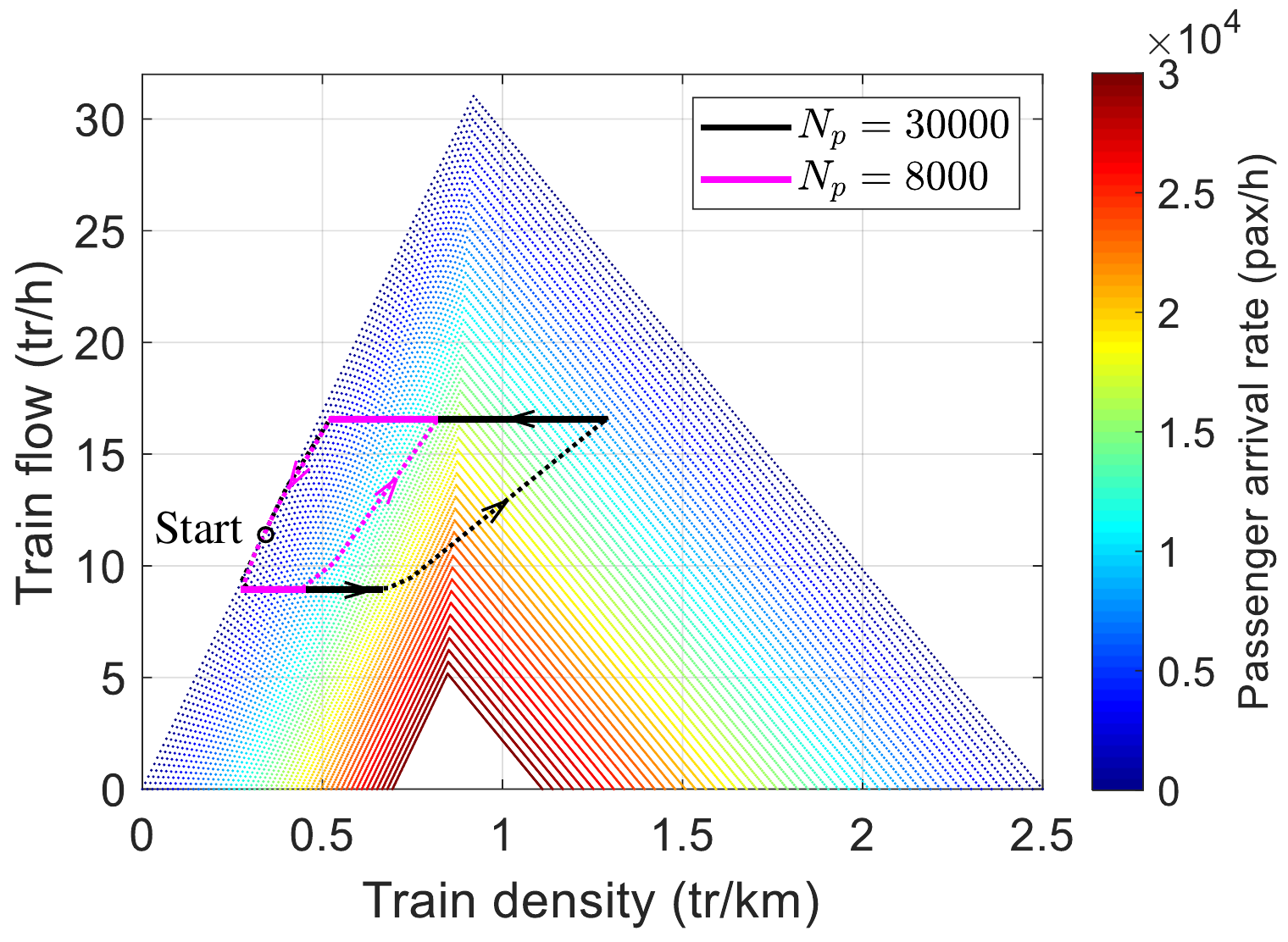}
    \caption{Dynamics of density and flow on train-FD.}
    \label{fig-dynaonfd}
\end{figure}

As discussed in the previous subsection, depending on the total travel demand $N_p$, the passenger arrival rate may have one or two peaks in FCF.
This phenomenon can be confirmed from the time evolution of the passenger arrival rate, $\mathrm{d}A_{p}(t)/\mathrm{d}t$, in Fig. \ref{fig-traindynamics}(b); a larger peak occurs close to the arrival time of passengers departing from the system just before $t_{m}$, and the other peak occurs near the end of rush hour. 
The mechanism behind this observation can be understood from the evolution of $k(n)$ and $q(n)$ on train-FD in Fig. \ref{fig-dynaonfd}.
The black line exhibits that the evolution of $(k(n), q(n))$ for the demand $N_p=30,000$ starts from the left boundary of the train-FD and moves along a counter-clockwise closed loop during rush hour. 
The dotted line indicates the sudden change in traffic states because of the discontinuity of travel time derivatives at $t_0$, $t^*$, and $t_{ed}$.
The lower part of the loop ($q(n)<12$~tr/h) represents the dynamics of trains departing from the system during [$t_0, t^*$], whereas the upper part represents the dynamics during [$t^*, t_{ed}$]. 
The maximum of $a_{p}(t)$ is reached at the lower-right corner of the loop, whereas the other peak occurs at the critical density in the upper part of the loop.
This two-peak phenomenon is a notable novel characteristic of the equilibrium distribution of passenger arrivals for a congested rail transit system, which should be verified through empirical observations.


However, when the total travel demand is rather low and all trains operate in the free-flow regime of train-FD, which corresponds to the FF mentioned in the previous subsection, $a_{p}(t)$ is maximized when the on-time train enters the system.
The magenta line in Fig. \ref{fig-dynaonfd} depicts the evolution of $(k(n), q(n))$ for this low-demand case.

\begin{figure}[t]
    \centering
    \subfigure[Time values for FCF]{\includegraphics[scale=0.34]{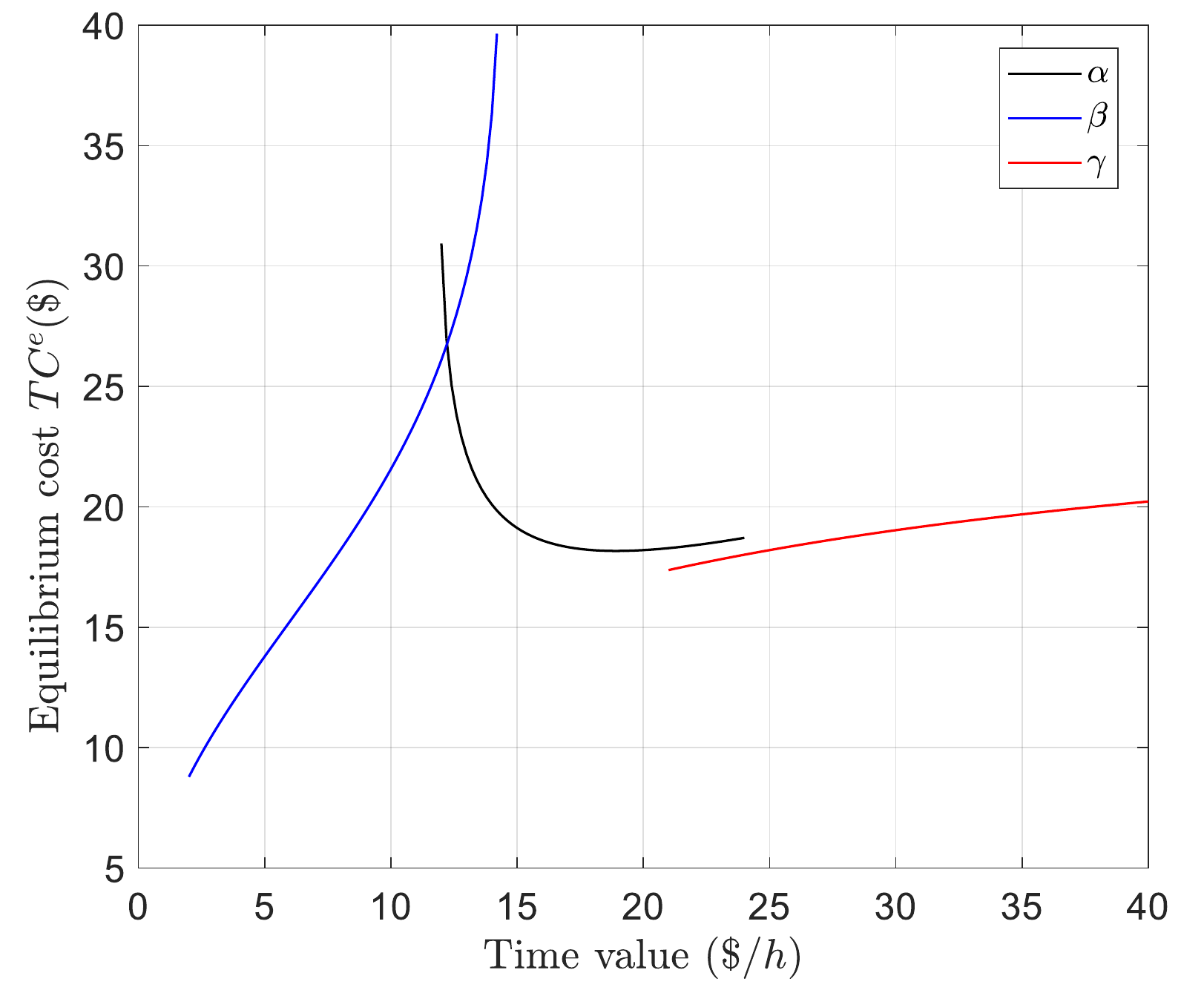}}\hspace{5pt}
    \subfigure[$N_{p}$ under different train inflow settings]{\includegraphics[scale=0.26]{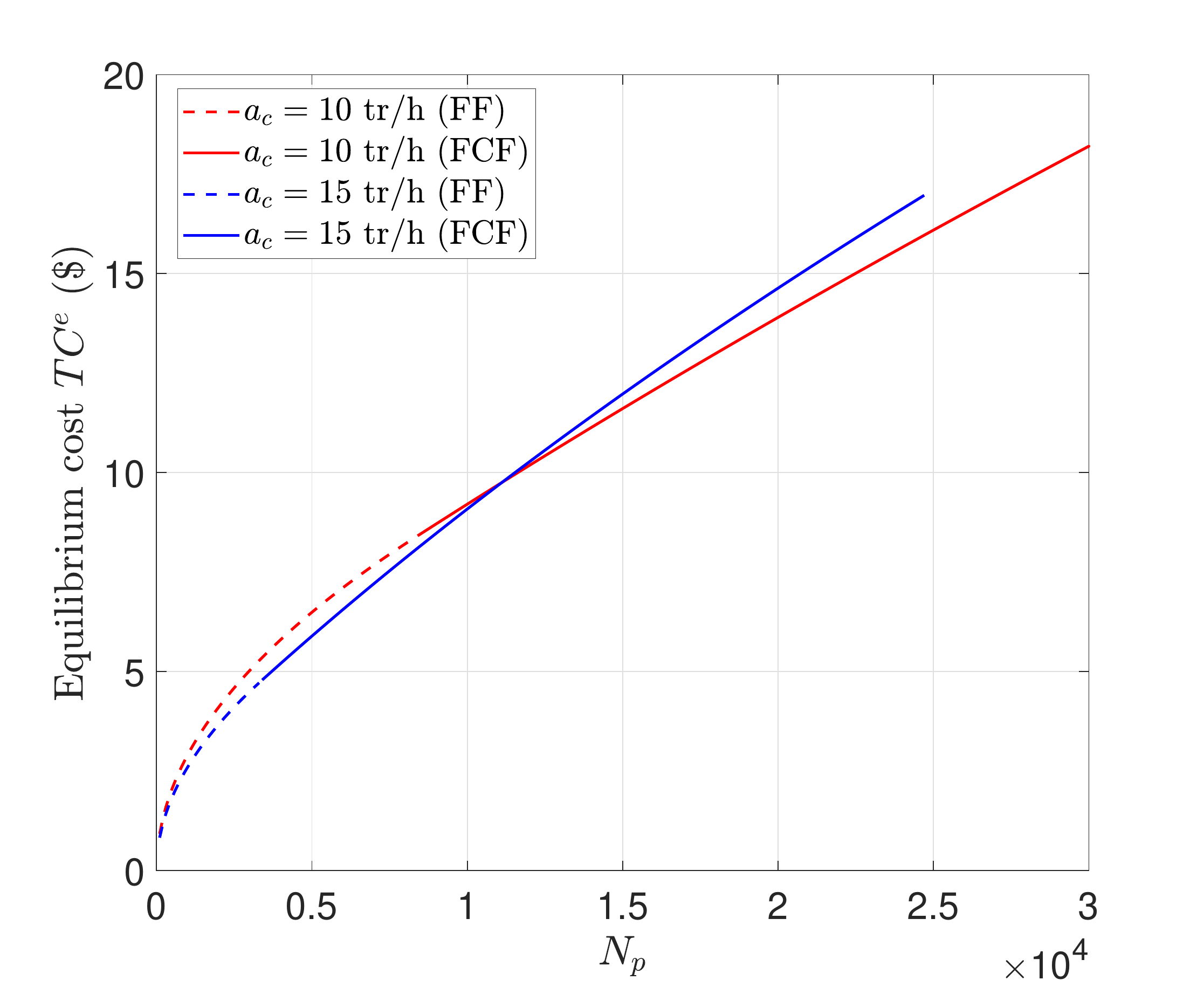}}
    \caption{Relations between equilibrium cost and other parameters.} 
    \label{fig-sensi}
\end{figure}

Next, we numerically evaluate the sensitivities of the equilibrium cost $TC^e$ to the time values for FCF.
Given the basic setting of time values in Table \ref{t-pasetting} and generally $\gamma>\alpha>\beta$, we set the test ranges of time values as $\alpha\in[9,24]~\$/h$, $\beta\in[2,18]~\$/h$, and $\gamma\in[21,40]~\$/h$.
When one time value is tested, all other parameters take the values listed in Table \ref{t-pasetting}.
Additionally, the parameter settings leading to $R^2-4U(S-N_{p})<0$ are eliminated.
The sensitivity test results are presented in Fig. \ref{fig-sensi}(a).
Although $TC^e$ increases with the increase in both $\beta$ and $\gamma$, it is more sensitive to the increase in $\beta$, especially when $\alpha-\beta$ approaches 0.
Besides, $TC^e$ first decreases with an increase of $\alpha$ but subsequently slowly increases with an increase in $\alpha$.

Finally, we present the relationship between $N_{p}$ and $TC^e$ under the two train inflow settings in Fig. \ref{fig-sensi}(b), which reveals that $TC^e$ monotonically increases with an increase in $N_{p}$ in both FF and FCF.
When trains are dispatched with a higher frequency, on-track congestion easily occurs; thus, FCF starts from a smaller $N_{p}$ for $a_c=15$ tr/h.
When $N_{p}$ is small (e.g., $N_{p}<1\times 10^4$), adopting a higher constant inflow can reduce the equilibrium cost even if some of the trains operate in the congested regime.
However, when $N_{p}$ is sufficiently large (e.g., $N_{p}>2\times 10^4$), the high constant inflow conversely leads to a higher equilibrium cost because $TC^{e}$ grows faster with the increase in $N_{p}$ under the higher inflow condition in FCF.
This result implies that adopting a timetable that dispatches trains as frequently as possible may not be always be appropriate from the perspective of reducing the equilibrium cost when passenger demand is high.

\section{A simple time-dependent timetable optimization} \label{s5}
This section presents the optimization problem of a time-dependent timetable pattern as an application of the proposed model.
The first subsection describes the problem setting, and the second subsection presents the results and provides their interpretations.

\subsection{Problem setting}
\begin{figure}[t]
    \centering
    \includegraphics[width=0.5\textwidth]{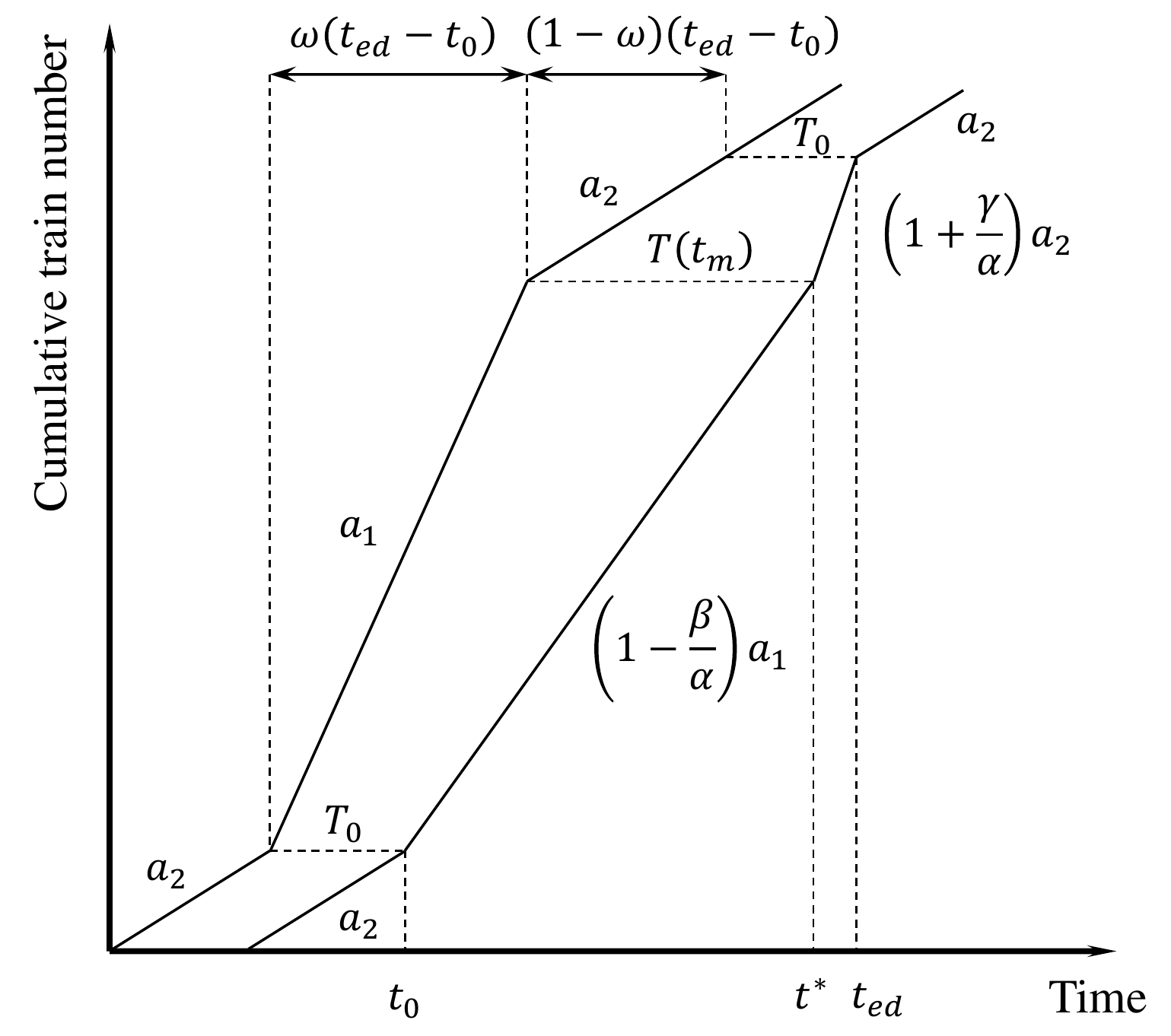}
    \caption{A simple time-dependent timetable pattern.}
    \label{fig-TBopticum}
\end{figure}

We consider the following simple time-dependent timetable pattern:  
\begin{enumerate}
    \item Two dispatch frequencies (or train inflows) $a_1$ and $a_{2} \ (a_{1}\ge a_2)$ are used.
    \item The train inflow is $a_{2}$ initially; it becomes $a_{1}$ from the beginning of the rush hour and lasts until the time at which the train carries the passenger departing from the system at $t^{*}$ on time, and then back to $a_{2}$, as displayed in Fig. \ref{fig-TBopticum}. 
\end{enumerate}
The first condition is widely adopted in practice. 
The second condition may be necessary to avoid degrading the train system considerably under user equilibrium (i.e., the timings of the inflow changes are expected to be near-optimum). 
Specifically, this condition is aimed to prevent train outflow from becoming very low in the first half of the rush hour, and the traffic state from entering the congested regime in the second half.
According to Fig. \ref{fig-TBopticum}, the ratio $\omega \in (0,1)$ of the duration for $a_{1}$ to the rush hour is given as a constant:  
\begin{align}
    &\omega = \frac{\gamma (\alpha - \beta)}{\alpha (\beta + \gamma)}
\end{align}

Thanks to the second condition that exploits the characteristic of the equilibrium, the optimization problem of determining $(a_{1}, a_{2})$ to minimize passengers' travel cost can be expressed as the following concise mathematical problem with equilibrium constraints (MPEC).
\begin{align} \label{eq-TBoptmobj}
    & \min_{a_1 \geq a_2 > 0} \ TC^e\left(a_1,a_2 \mid N_{p}\right) \\ 
    & \text{subject to} \quad \omega a_1 + (1-\omega)a_2 \leq a_0 \label{eq-TBoptmcon}
\end{align}
where $TC^e\left(a_1,a_2 \mid N_{p}\right)$ is the equilibrium travel cost as a function of the decision variables. 
Eq. \eqref{eq-TBoptmcon} indicates the dispatch capacity constraint, where $a_{0}$ is the maximum available train inflow during rush hour. 
This problem can be easily solved by a brute-force search using Algorithm 1. 
Because the train inflow is time-dependent, the condition $a_{p}(n)\ge0$ should be checked while evaluating the objective function.

\subsection{Results}
Before moving to the numerical results, an analytical discussion of this simplified timetable optimization problem in FF is made.
We expect the analytical solution of the optimization problem in FF to be near-optimum because operation in the congested regime vainly increases the travel time cost.
Following a process similar to that introduced in Section \ref{s4-1}, the equilibrium cost $TC^{e}$ in Eq. \eqref{eq-TBoptmobj} in FF can be obtained explicitly as follows:
\begin{align} \label{eq-TBTC_ff}
    TC^{e}\left(a_1,a_2 \mid N_{p}\right) =  \sqrt{\frac{2\alpha LN_{p}}{\mu l\left[\left(\frac{1}{\beta}-\frac{1}{\alpha}\right)a_1+\left(\frac{1}{\gamma}+\frac{1}{\alpha}\right)a_2\right]}}.
\end{align}
If the dispatch capacity constraint (i.e., Eq. \eqref{eq-TBoptmcon}) is inactive, it can be proved (see Appendix \ref{appendix}) that $TC^{e}$ is minimized in FF when the following expression holds:
\begin{align} \label{eq-TBa1a2ratio}
    \frac{a_1}{a_2} = \frac{\zeta_2}{\zeta_1} \quad \Leftrightarrow \quad \frac{a_1}{a_2}=\frac{(\alpha+\gamma)(2\alpha-\beta)}{(\alpha-\beta)(2\alpha+\gamma)}.
\end{align}
It is seen that $a_1/a_2$ should increase with a decrease of $\alpha$ and with an increase of $\beta$ or $\gamma$.
Besides, by referring to Eq. \eqref{qndiscuss}, in FF, average train flow $q(n)$ remains unchanged during the entire equilibrium period because $q(n)=\zeta_1a_1=\zeta_2a_2$.  

\begin{figure}[t]
    \centering
    \includegraphics[width=0.6\textwidth]{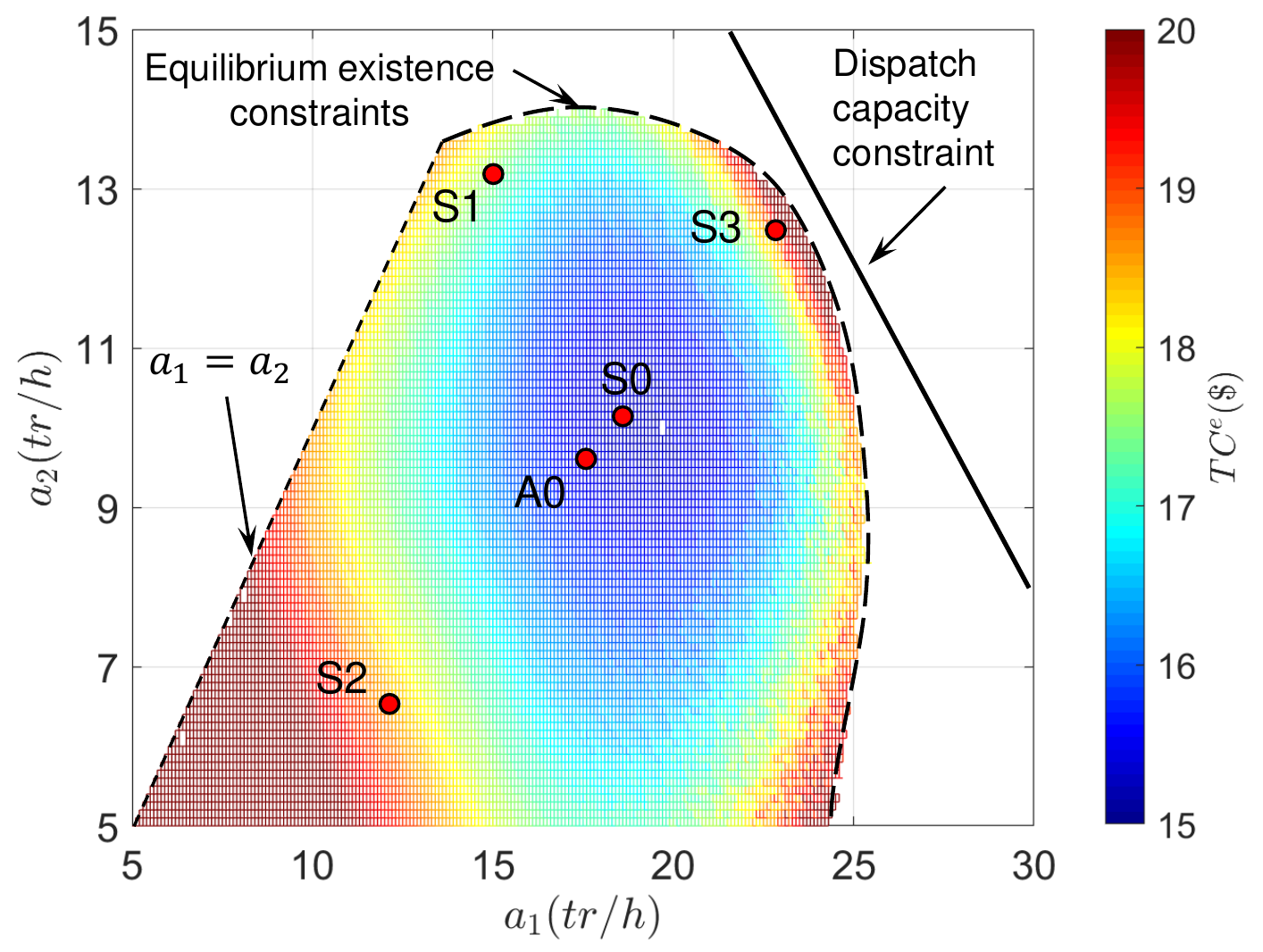}
    \caption{Counter plot of the objective function.}
    \label{fig-TBresult}
\end{figure}

Next, the numerical optimization results under the parameter settings in Table \ref{t-pasetting} (and $a_{0}=18$ tr/h) are displayed for Case WT1 in Fig.~\ref{fig-TBresult}. 
The horizontal and vertical axes represent the high inflow rate $a_{1}$ and low inflow rate $a_{2}$, respectively. 
Color represents the value of the objective function.
We evaluated the objective function every $0.1$~tr/h for both inflow rates. 

\begin{figure}[t]
\centering
\subfigure[Cumulative number of trains (double arrow line: the equilibrium rush period)]{\includegraphics[scale=0.37]{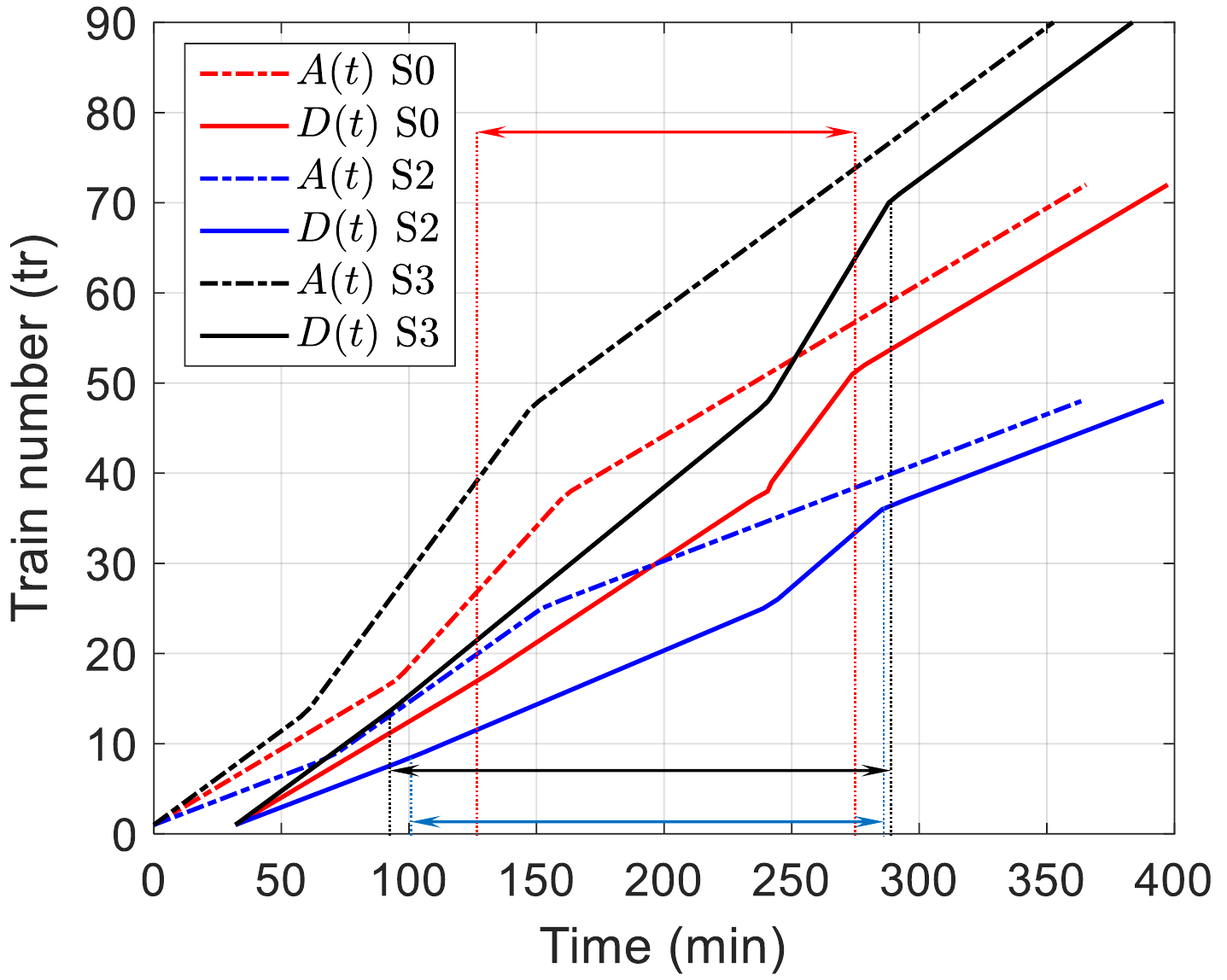}}\hspace{5pt}
\subfigure[Density and flow on train-FD]{\includegraphics[scale=0.37]{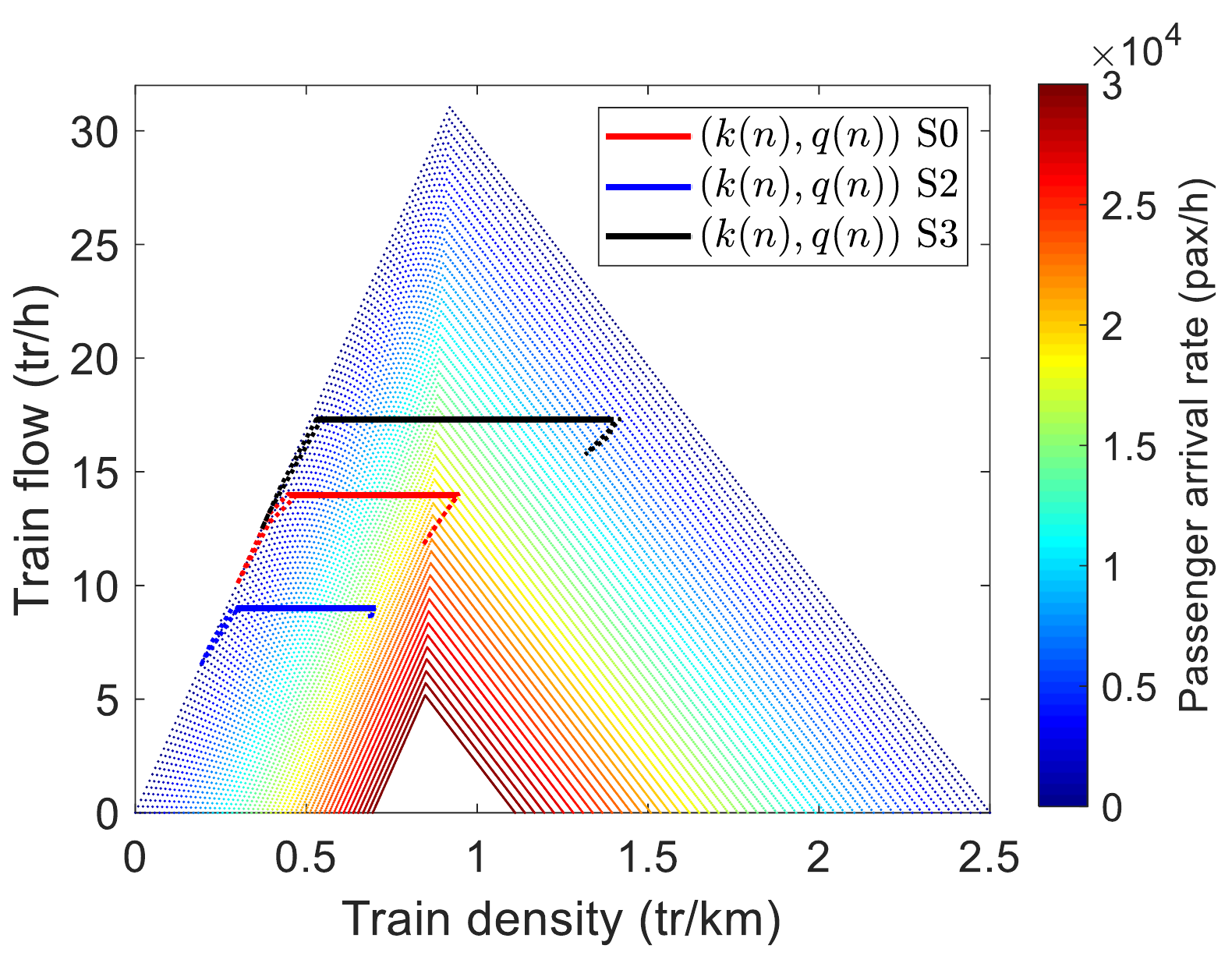}}
\caption{Dynamics of the rail transit system under time-dependent timetable patterns.} \label{fig-TBresultdetail}
\end{figure}

This figure reveals that the objective function is almost convex, and a unique optimal solution (S0) is obtained.
The analytical solution of this optimization problem in FF is plotted as A0 on Fig. \ref{fig-TBresult} ($a_1 = 17.7$~tr/h, $a_2 = 9.6$~tr/h, and $TC^{e}=15.54$~\$ from numerical results).
This $TC^{e}$ is close to that in scenario S0 ($TC^{e}=15.14~\$$), which indicates that the solution in FF is near-optimum.
The numerical optimization results also indicate that maximizing the dispatch frequency can increase the equilibrium cost.
This phenomenon is a particular type of ``capacity increasing paradox", known to occur in equilibrium transportation systems \citep[e.g.,][]{braess1968,arnott1993paradox}.
To understand the reason for this phenomenon, the train dynamics for scenarios S0, S2, and S3 are displayed in Fig. \ref{fig-TBresultdetail}. 
The ratio $a_{1}/a_{2}$ of the S2 and S3 scenarios is the same as the optimal scenario S0, but with different average inflow rates. 
From the train cumulative curves in Fig. \ref{fig-TBresultdetail}(a), it can be seen that the equilibrium rush period for S0 is shorter than that of S2 and S3. 
The reason for this result can be understood from Fig. \ref{fig-TBresultdetail}(b).
For S0, a high passenger arrival rate was achieved while maintaining an appropriately high train flow. 
This result indicates that, at the optimum, the intention of the second condition in the previous subsection is achieved. 
By contrast, for S2 and S3, a train flow that is either too low or too high cannot accommodate the high passenger arrival rate.
For scenario S2, passenger arrival rate in Eq. \eqref{apn_ff} is considerably reduced because of the insufficient dispatch frequency (or low $a(t)$).
As a result, longer rush period is required to carry the same number of passenger $N_{p}$.
For scenario S3, a similar conclusion is obtained because a considerable proportion of trains operate in the congested regime of the train-FD (large $a(t)$ in Eq. \eqref{apn_cong} results in small $a_{p}(n)$).

In addition, the average train flow $q(n)$ still maintains at an almost constant level during the equilibrium period under the optimal setting S0, which indicates that $a_1/a_2$ also follows Eq. \eqref{eq-TBa1a2ratio} even if S0 has a small number of trains operated in the congested regime.
This observation may imply that flattening the train operation performance could be a useful strategy to reduce the equilibrium cost in a more general case.

Finally, Table \ref{t-TBcompare} summarizes the travel costs for the timetable settings S0--S3 in Fig. \ref{fig-TBresult}.
Scenario S1 represents the case with the same average inflow as S0, but a smaller difference between $a_1$ and $a_2$. 
We can observe that the total travel cost $\sum TC$ ($\sum TC=TC^eN_{p}$) in scenarios S1--S3 are significantly higher than that of optimal scenario S0.
Specifically, by comparing S0 and S1, the increase in the total schedule delay cost $\sum SDC$  (31\%) is greater than that of the total travel delay cost $\sum TDC$ (5\%).
This suggests a primary deficiency of timetable patterns with a low $a_1/a_2$ ratio is that passengers cannot arrive at their desired arrival time $t^{*}$ sufficiently.
A similar property can be obtained for a scenario with an redundant train supply (S3). 
However, when the train supply is insufficient (S2), passengers would suffer from a significantly longer travel delay ($\sum TDC$ increases by 27\% compared with S0). 

\begin{table}[t]
  \centering
  \caption{Comparison of travel costs for different timetable patterns.}
    {\begin{tabular}{cccccccc}
    \toprule
     & \multirow{2}*{\shortstack{$a_1$\\(tr/h)}} & \multirow{2}*{\shortstack{$a_2$\\(tr/h)}} &
    \multirow{2}*{\shortstack{average\\inflow (tr/h)}} &
    \multirow{2}*{\shortstack{$\sum TDC$\\($10^4~\$$)}} & \multirow{2}*{\shortstack{$\sum SDC$\\($10^4~\$$)}} & \multirow{2}*{\shortstack{$\sum TC$\\($10^4~\$$)}} & \multirow{2}*{\shortstack{$TC$ change\\(\%)}}\\
     & & & & & & & \\
    \midrule
    S0 & 18.7 & 10.1 & 14.0 & 29.30 & 16.13 & 45.43 & - \\
    S1 & 15.0 & 13.2 & 14.0 & 30.68 & 21.11 & 51.79 & +14.0 \\
    S2 & 12.0 & 6.5 & 9.0 & 37.30 & 18.46 & 55.76 & +22.7\\
    S3 & 23.1 & 12.5 & 16.5 & 29.75 & 29.59 & 59.34 & +30.6\\
    \bottomrule
    \end{tabular}}
  \label{t-TBcompare}
\end{table}

\section{Conclusions} \label{s6}
This study proposed a macroscopic model to describe the equilibrium distribution of passenger arrivals for the morning commute problem in a congested urban rail transit system.
We first developed a model for the morning commute problem in rail transit based on the train-FD and derived the equilibrium conditions. 
Next, we proposed a solution method and examined the characteristics of the equilibrium through both analytical discussion and numerical examples. 
Finally, by applying the proposed model, we analyzed a simple time-dependent timetable optimization problem with equilibrium constraints.

The proposed model is not only mathematically tractable but can also thoroughly consider the relations among passenger concentration, on-track congestion, and time-dependent timetable in a congested rail transit system. 
This enables us to investigate the characteristics of the equilibrium and optimal design of the timetable in a simple manner.
The contributions of this study are summarized as follows: (i) we revealed that the evolution of rail transit flow and density under equilibrium; (ii) we obtained the closed-form solutions of the equilibrium in two of three patterns when train dispatch frequency is constant; we further found that, under equilibrium,   
(iii) a ``capacity increasing paradox" exists in which a higher dispatch frequency can increase the equilibrium cost, 
(iv) an insufficient supply of rail transit mainly increases the travel delay cost while redundant supply increases the schedule delay cost, and  
(v) the average train flow maintains at an almost constant level under an optimal timetable setting.

The straightforward extensions of the proposed model include the consideration of elastic demands and captive users \citep[e.g.,][]{gonzales2012}.
For the elastic demands, we only need to specify the travel demand $N_{p}(TC^e)$ as a monotonically decreasing function of the equilibrium travel cost \citep[e.g.,][]{arnott1993elastic,zhou2005}.
Including captive users can be achieved by modifying $a_{p}(n)$ in Eq. (\ref{eq-ap}) as $a_{p}(n) = a_{pc} + a_{pf}(n)$, where $a_{pc}$ is the arrival rate of captive users, and $a_{pf}(n)$ is the arrival rate of flexible users for train $n$.

In this study, rail transit is assumed to be a homogeneous system in which both stations and passenger demand are evenly distributed. 
Thus, a train-FD model applicable to a heterogeneous railway system should be developed to deal with a more realistic situation.  
Considerations of heterogeneity in passenger preferences (i.e., the value of time) \citep{newell1987,akamatsu2020} and the costs/revenue of the transit agency in the optimization of timetable/fare settings are also important topics. 
The design of pricing schemes could be another fruitful future work.
By using the proposed model, we could obtain insights into not only the first-best pricing scheme but also the second-best schemes \citep[e.g., step tolls in][]{arnott1990economics,laih1994,lindsey2012}, which are generally formulated as MPEC. 

\textbf{Author contribution} Conceptualization: Kentaro Wada; Methodology: Kentaro Wada and Jiahua Zhang; Formal analysis and investigation: Jiahua Zhang; Writing - original draft preparation: Jiahua Zhang; Writing - review and editing: Takashi Oguchi and Kentaro Wada; Funding acquisition: Kentaro Wada; Supervision: Takashi Oguchi and Kentaro Wada.


\section*{Acknowledgements}
This study was financially supported by JSPS KAKENHI Grant No. JP17H03320.

\begin{appendices}
\section{}\label{appendix}
To simplify the discussion, the available train inflow is assumed to be sufficient, which indicates that constraint Eq. \eqref{eq-TBoptmcon} is inactive.
To minimize $TC^{e}$ under a given $N_{p}$, $a_1$ and $a_2$ should be as large as possible, according to Eq. \eqref{eq-TBTC_ff}.
Meanwhile, to ensure free-flow operation, another two constraints should be added, similar to Eq. \eqref{ffkas}:
\begin{align}
    \left\{
    \begin{aligned}
    \frac{\zeta_1}{L}a_{1}\left[T_0 + \frac{\beta}{\alpha}(t_{m}-t_0)\right]\leq \frac{1}{l}\left[1+\zeta_1 a_{1}\left(\frac{l-\delta}{v_{f}}-\tau\right)\right],\\
    \frac{\zeta_2}{L}a_{2}\left[T_0 + \frac{\beta}{\alpha}(t_{m}-t_0)\right]\leq \frac{1}{l}\left[1+\zeta_2 a_{2}\left(\frac{l-\delta}{v_{f}}-\tau\right)\right].\\
    \end{aligned}
    \right.
\end{align}
Substituting $TC^{e}=\beta(t_{m}-t_{0})$ into these two constraints, we obtain the following expression:
\begin{align} \label{eq-TBFFcons}
    \left\{
    \begin{aligned}
    G_{1}\left(a_1,a_2 \mid N_{p}\right) = \zeta_1 a_1\left[\frac{l}{L}\left(T_{0}+\frac{TC^{e}}{\alpha}\right)-\frac{l-\delta}{v_{f}}+\tau\right]\leq 1,\\
    G_{2}\left(a_1,a_2 \mid N_{p}\right) = \zeta_2 a_2\left[\frac{l}{L}\left(T_{0}+\frac{TC^{e}}{\alpha}\right)-\frac{l-\delta}{v_{f}}+\tau\right]\leq 1.\\
    \end{aligned}
    \right.
\end{align}

Next, it becomes easy to prove that $a_1$ and $a_2$ are maximized when $G_1=1$ and $G_2=1$ hold simultaneously, which also indicates that $\zeta_1a_1=\zeta_2a_2$.
This is because if any of this two inequalities is smaller than 1, both $a_1$ and $a_2$ can be larger while not violating this two constraints.
For instance, selecting an appropriate large $a_1$ and $a_2$ that make $G_1=1$ but $G_2<1$.
Then, we can raise $a_2$ until $G_2=1$ while maintaining $a_1$ unchanged because $\partial G_2/\partial a_2>0$.
Meanwhile, because $\partial G_1/\partial a_2<0$, increasing $a_2$ leads to $G_1<1$, which indicates that $a_1$ can also be increased again.
This iteration of increasing $a_1$ and $a_2$ stops only when $G_1=1$ and $G_2=1$ at the same time.
On one hand, $\partial G_1/\partial a_2<0$ and $\partial G_2/\partial a_1<0$ are obvious because $a_2$ only appears in the denominator of $TC^{e}$ in $G_1$; similarly $a_1$ only appears in the denominator of $TC^{e}$ in $G_2$.
On the other hand, we can prove that when $a_1>0$ and $a_2>0$, $\partial G_1/\partial a_1>0$, and $\partial G_2/\partial a_2>0$.
For example, $\partial G_1/\partial a_1$ can be derived as follows:
\begin{align}
    \partial G_1/\partial a_1 = & \frac{TC^{e}l}{\alpha L}\zeta_1\left[1 - \frac{1}{2}\frac{\left(\frac{\alpha}{\beta}-1\right)a_1}{\left(\frac{\alpha}{\beta}-1\right)a_1 + \left(\frac{\alpha}{\gamma}+1\right)a_2}\right]\\ \notag
    & + \zeta_1\left(t_{b0} + \delta/v_{f} + \tau\right) > 0.
\end{align}
$\partial G_2/\partial a_2$ can be derived similarly, and it is also larger than zero.

\end{appendices}

\bibliography{Train_DTCE}

\end{document}